%% file: manuscript.tex
\documentclass[referee,pdflatex,sn-apa]{sn-jnl}

\usepackage{graphicx}%
\usepackage{multirow}%
\usepackage{amsmath,amssymb,amsfonts}%
\usepackage{amsthm}%
\usepackage{mathrsfs}%
\usepackage[title]{appendix}%
\usepackage{xcolor}%
\usepackage{textcomp}%
\usepackage{manyfoot}%
\usepackage{booktabs}%
\usepackage{algorithm}%
\usepackage{algorithmicx}%
\usepackage{algpseudocode}%
\usepackage{listings}%
\usepackage{tabularx}
\usepackage{rotating}

\theoremstyle{thmstyleone}%
\theoremstyle{thmstyletwo}%

\theoremstyle{thmstylethree}%

\makeatletter
\AtBeginDocument{%
  \def\ps@headings{%
    \let\@oddfoot\@empty\let\@evenfoot\@empty
    \def\@oddhead{\hfill\headerfont\thepage}%
    \let\@evenhead\@oddhead
    \let\@mkboth\markboth}%
  \def\ps@titlepage{%
    \let\@oddfoot\@empty\let\@evenfoot\@empty
    \def\@oddhead{\hfill\headerfont\thepage}%
    \let\@evenhead\@oddhead}%
  \pagestyle{headings}%
}
\makeatother

\begin{document}

\title[Handwriting-Process Profiling Instrument]{Profiling Handwriting-Process Deviations in Developmental Dysgraphia: An Open, Normatively-Referenced Instrument}


\author[1]{\fnm{Jihyun} \sur{Mun}}\email{jihyun.mun@telecom-sudparis.eu}

\author*[1]{\fnm{Mounîm A.} \sur{El-Yacoubi}}\email{mounim.el\_yacoubi@telecom-sudparis.eu}

\affil*[1]{\orgdiv{Samovar}, \orgname{T\'{e}l\'{e}com SudParis, Institut Polytechnique de Paris}, \orgaddress{\street{19 Place Marguerite Perey}, \city{Palaiseau}, \postcode{91120}, \country{France}}}

\presentaddresstxt{Corresponding author telephone}
\presentaddress{+33 6 83 70 12 89}

\input{sections/0_abstract}

\maketitle

\input{sections/1_introduction}
\input{sections/2_methods}
\input{sections/3_results}

\input{sections/4_discussion}
\input{sections/5_conclusion}

\backmatter
\input{sections/back_matter}

\clearpage
\bibliography{sn-bibliography}

\input{sections/appendix}

\end{document}

%% file: sections/0_abstract.tex

\abstract{%
\linespread{1.667}\selectfont%
Online handwriting captured on digitizing tablets yields hundreds of kinematic, temporal, spatial, and dynamic features.
These features are rarely organized into interpretable, reusable constructs, and, although standardized tests norm the handwritten product, no openly available instrument places a child's handwriting process relative to a verified-typical reference.
We introduce and validate an open measurement instrument for profiling handwriting-process deviations in children with developmental dysgraphia from sentence-level online handwriting.
The instrument comprises (i)~a literature-only vocabulary, fixed before any cohort analysis, organizing 136 online-handwriting features into 12 handwriting-process domains, and (ii)~an age- and sex-adjusted normative reference framework, fit on verified-typical children only, that turns those features into a 12-axis deviation profile with per-child bootstrap uncertainty.
We establish its measurement properties on the DiaGraMo cohort (N=257 Czech children aged 8--12; 110 verified-typical, 147 with dysgraphia): the vocabulary is structurally coherent, and the reference is calibrated, parsimonious, and leakage-free.
On the same cohort, five of twelve domains separate the groups at BH~$q<.05$ (Cliff's~$\delta$ +0.19 to +0.51), on spatial, temporal, and pen-orientation processes, and 96.5\% of participant~$\times$~domain scores have a bootstrap CI narrower than one z-unit.
No classifier is trained: the instrument reports uncertainty-aware deviation scores rather than a diagnostic label, and its outlier rate is not a diagnostic rate.
The vocabulary, the analysis code, and an open reference implementation that scores new participants are all openly released, giving handwriting and dysgraphia researchers a reusable, validated instrument for situating individual children against a normative reference.
}

\keywords{Handwriting kinematics, Online handwriting, Measurement instrument, Normative reference, Deviation profiling, Developmental dysgraphia}


%% file: sections/1_introduction.tex
\section{Introduction}
\label{sec:intro}

Handwriting is a foundational transcription skill \citep{maldarelli2015development} and a complex perceptual-motor task \citep{danna2023tools}.
Producing legible script is multi-dimensional, demanding fine motor control, attention, working memory for letterforms, and visuomotor coordination bound to orthographic rules \citep{maldarelli2015development}, and staged, running from lexical and orthographic preparation through to motor execution \citep{xu2026stroke}.
Handwriting acquisition is gradual.
Formal instruction begins around age five and mastery takes roughly a decade; automaticity is only partial at age ten and near-complete by fourteen, with quality maturing before speed.
As handwriting automatizes through the middle-school years it demands less attention, leaving more cognitive capacity for planning and composing text \citep{feder2007handwriting, danna2023tools}.

For a substantial minority of children, handwriting nonetheless fails to develop as expected.
This condition, developmental dysgraphia, is difficulty acquiring writing skills despite adequate learning opportunities, sufficient cognitive potential, and the absence of neuropathology, neurological problems, or gross sensory-motor dysfunction \citep{chung2020disorder, mccloskey2017developmental, rosenblum2004handwriting}, with writing performance below that expected for the child's grade level \citep{dohla2016developmental}.
Reported prevalence ranges from roughly 7\% to 30\% of school-aged children \citep{dohla2016developmental, chung2020disorder}, and it is more common in boys than in girls \citep{dohla2016developmental, chung2020disorder, danna2023tools}.
Its causes are multiple.
Proximal causes are failures to acquire components of the adult spelling system, such as orthographic long-term and working memory and sublexical sound-to-spelling conversion; distal causes are broader deficits in phonological processing, vision, or motor control.
Different children reach the same diagnostic label by different routes, so the condition is heterogeneous \citep{han2025motor}, and children who share the label are impaired in different sub-processes \citep{danna2023tools}.

Handwriting underlies so much of schooling that persistent difficulty constrains written expression and academic achievement \citep{feder2007handwriting, danna2023tools}.
Timely, well-targeted intervention can mitigate these consequences, but it presupposes accurate assessment: not only identifying which children struggle, but characterizing how each child's handwriting is impaired, so that support can be directed at the responsible processes.

\subsection{Assessing the handwriting process}
Several standardized tools assess handwriting, including the BHK \citep{hamstra1987beknopte} and the DASH-2 \citep{barnett2007detailed}, alongside other instruments in common use \citep{danna2023tools}.
All of them, however, share several limitations.
They are product-based: they score or rate the handwritten product (legibility, letter size, and tilt, with speed estimated as characters per unit time) and do not decompose the writing process into its constituent dimensions.
Because they read only the finished product, the motor, temporal, and dynamic operations that generated it remain concealed.
They are also subject to rater bias and to extraneous influences, and, being based on human rating, are time-consuming and demanding of human resources \citep{kunhoth2023exploration}.

Increasingly, handwriting is instead captured by digitizing tablets that record the pen trajectory at high sampling rates, registering pen position, pressure, tilt and azimuth, and on-surface versus in-air status.
From these signals a rich set of kinematic, temporal, spatial, and dynamic features can be derived that go beyond the static product.
The writing process, and not only its product, thereby becomes measurable.

A large body of work has sought to detect dysgraphia from this online handwriting signal.
Most studies take a machine-learning approach, extracting a handcrafted feature set that spans the kinematic, temporal, spatial, and dynamic aspects of handwriting and feeding it to a classifier, most often for binary classification \citep{drotar2020dysgraphia, rosenblum2016identifying, asselborn2018automated, kunhoth2024automated, kunhoth2023exploration, mekyska2016identification}.
Deep-learning approaches instead take the raw signal as input and train a model for the same binary task \citep{kunhoth2023cnn, manimekala2025dysgraphia, lomurno2023deep, bublin2022automated}.
These methods frequently report high classification accuracy, yet, from a measurement standpoint, they share two limitations.
First, their features are not organized into interpretable, named constructs: consumed as a flat vector with no correspondence to handwriting-process dimensions, they form no reusable, clinically interpretable vocabulary.
Second, the output is a single dysgraphic/typical label rather than a per-dimension score, so it cannot indicate which handwriting process is impaired.
Post-hoc attribution methods do not fill this gap: they return feature importances for the model's decision, not construct-level measures of where a child stands, and those importances can change when the model is retrained in other settings.

\subsection{Normative referencing of individual deviation}
Because handwriting develops with age, a raw score means different things at different ages, so judging atypicality requires a demographically referenced norm.
The standardized instruments already provide one for the product, with both the BHK and the DASH-2 situating a child against grade- or age-based normative data \citep{hamstra1987beknopte, barnett2007detailed}.
No such reference exists for the process measures, and the detection methods above do not supply one, returning a label rather than a graded position relative to healthy, age- and sex-matched peers.
Normative modeling \citep{marquand2016understanding} formalizes this kind of referencing: it maps the full range of population variation and treats an individual's deviation as an extreme value within a distribution learned from a reference cohort, supporting probabilistic statements about which participants deviate and by how much.
The closest prior handwriting profiling, proposed by \citet{asselborn2020extending}, applied age-based $z$-scoring of features.
Its reference was drawn from all participants rather than from healthy children alone, which blurs the boundary of normal development; it operated on individual features rather than on process constructs; and it reported no formal validity evidence.

Taken together, these gaps mean that no openly available, validated, verified-typical-referenced, and reusable instrument exists for handwriting-process profiling.
This motivates a reframing of the task, from ``is this child dysgraphic?'' to ``which handwriting-process dimensions deviate from healthy, age-matched norms?''

\subsection{The present study}
We introduce and validate an open measurement instrument for profiling handwriting-process deviations in children with developmental dysgraphia, from sentence-level online handwriting.

The instrument has two components.
The first is a literature-based vocabulary that organizes 136 online-handwriting features, spanning kinematic, temporal, spatial, and dynamic aspects, into twelve interpretable handwriting-process domains grounded in the motor-control literature.
This feature-to-domain mapping was frozen before any analysis of the cohort.
The second is an age- and sex-adjusted normative reference framework.
We fit a linear model of age and sex on verified-typical children only, compute residuals for all children, $z$-score those residuals against the healthy residual standard deviation, and average them within each domain to give a per-child, twelve-axis deviation profile with bootstrap uncertainty.
Leave-one-subject-out refitting keeps the inference leakage-free.

We establish the instrument's measurement properties on the DiaGraMo cohort (N${=}$257 Czech children aged 8--12; 110 verified-typical and 147 with developmental dysgraphia) \citep{zvonvcakova2026multimodal}.
We first examine the structural coherence of the twelve domains, then the calibration, parsimony, and leakage-free inference of the reference framework, and then cross-sectional clinical validity through known-groups separation between dysgraphic and verified-typical children.
We finally characterize individual-profile reliability and a hypothesis-generating clinical interpretation.
The instrument reports uncertainty-aware deviation scores rather than a diagnostic label, and the rate at which it flags deviations is not an estimate of how many children are dysgraphic.
As a supporting, practical-use layer, we provide an open reference implementation that computes the same twelve-axis profile for new participants.

The paper makes three contributions: a normative framework for developmental dysgraphia that profiles children against a verified-typical reference; a construct-level account of handwriting deviation; and an open, reusable implementation for individual profiling, released together with the vocabulary and the analysis code.

%% file: sections/2_methods.tex

\section{Methods}

\subsection{Dataset and tasks}

We use the DiaGraMo dataset \citep{zvonvcakova2026multimodal}, a multimodal corpus of online handwriting and cognitive-assessment data from 276 Czech primary-school children aged 8--12, of whom 161 were diagnosed with developmental dysgraphia.
The protocol comprises 16 tasks: graphomotor primitives (spiral, saw, loops, rainbow) and linguistic tasks (sentence copying and dictation).
Handwriting was recorded with a Wacom Cintiq 16 tablet and a digital pen, capturing pen position ($x$, $y$), pressure, and tilt at a mean sampling rate of 167~Hz.
Each child additionally completed standardized assessments of general cognitive ability (WJ-IV), visuospatial processing (RCFT), phonological awareness (BACH), and questionnaire-based handwriting proficiency (HPSQ-C).

Of the 276 children in the published cohort, 257 form the final sample (110 healthy, 147 dysgraphic).
We excluded children lacking a completed sentence-level linguistic protocol (TSK3$+$15 or TSK4$+$16), a known diagnostic group, or clean data-quality flags, together with three healthy children whose missing age precludes age-referencing.
Twenty-nine of the dysgraphic children also carry a dyslexia diagnosis, used only for the comorbidity stratification (Appendix~\ref{app:comorbid}).
Overall age is $9.78 \pm 0.73$~yr (range $8.20$--$11.92$); the sample is 73.2\% male (188 male / 69 female).
Participant characteristics, broken down by linguistic protocol and diagnostic group, are summarized in Table~\ref{tab:cohort}.

\begin{table}[t]
\centering
\caption{Participant characteristics of the final analyzed DiaGraMo cohort
($N=257$), by linguistic protocol and diagnostic group. Age in years
(mean~$\pm$~SD); \%F = female proportion. Protocol A = TSK3$+$15,
Protocol B = TSK4$+$16.}
\label{tab:cohort}
\small
\begin{tabular}{@{}cccccc@{}}
\toprule
Protocol & Group & $n$ & M / F & \%F & Age (mean~$\pm$~SD) \\
\midrule
A & Healthy    & 51  & 32 / 19 & 37.3 & $9.14 \pm 0.51$ \\
A & Dysgraphic & 56  & 37 / 19 & 33.9 & $9.20 \pm 0.59$ \\
\multicolumn{2}{c}{A subtotal} & 107 & 69 / 38 & 35.5 & $9.17 \pm 0.55$ \\
\midrule
B & Healthy    & 59  & 48 / 11 & 18.6 & $10.29 \pm 0.36$ \\
B & Dysgraphic & 91  & 71 / 20 & 22.0 & $10.17 \pm 0.54$ \\
\multicolumn{2}{c}{B subtotal} & 150 & 119 / 31 & 20.7 & $10.22 \pm 0.48$ \\
\midrule
\multicolumn{2}{c}{Total} & 257 & 188 / 69 & 26.8 & $9.78 \pm 0.73$ \\
\botrule
\end{tabular}
\end{table}

\subsubsection{Linguistic-only task restriction}
Of the available tasks, we analyze only the linguistic ones, sentence-level copying and dictation, for two clinical reasons.
First, the clinical definition of dysgraphia is writing-centric rather than motor-centric: contemporary criteria conceptualize it as a disorder of written production spanning legibility, spelling, writing rate, and composition, rather than of graphomotor execution per se \citep{chung2020disorder}.
Second, the standard clinical assessment instruments are themselves built on writing tasks rather than graphomotor primitives, relying on copying, dictation, and spontaneous sentence writing \citep{danna2023tools}.

\subsubsection{Protocol--age confound disclosure}
The linguistic tasks were administered under two age-stratified protocols (Protocol~A: TSK3 dictation + TSK15 copy; Protocol~B: TSK4 dictation + TSK16 copy; Table~\ref{tab:cohort}).
Because protocol assignment follows age by design, protocol is collinear with age in the pooled sample, and the two protocols also differ in their tasks.
A pooled group difference therefore cannot be cleanly attributed to dysgraphia rather than to protocol or age.
Within each protocol, however, the healthy and dysgraphic groups are age-matched and perform identical tasks.
We therefore retain the pooled analysis but defend it in two ways.
The age- and sex-adjusted reference framework removes, by construction, the age variance with which protocol is collinear.
A sensitivity suite then replicates the group contrasts within protocol and quantifies any residual protocol effect (Section~\ref{sec:validation-design}).

\subsection{Feature extraction and the frozen vocabulary}
We first define the deterministic feature pipeline that converts each recording into a fixed feature vector, then organize that vector into the frozen vocabulary of handwriting-process domains that the rest of the analysis builds on.

\subsubsection{Online handwriting feature pipeline}
Each linguistic recording is converted into a fixed vector of 136 handcrafted online-handwriting features, computed deterministically from the pen trajectory.
These descriptors are drawn from the established online-handwriting literature on dysgraphia assessment \citep{drotar2020dysgraphia, rosenblum2016identifying, asselborn2018automated, kunhoth2024automated, kunhoth2023exploration, mekyska2016identification}.

\emph{Summary-statistics scheme.}
A handwriting recording is a multichannel time series, so a single number cannot describe any one of its signals; we therefore summarize each derived signal or set of segment-level values with a common battery of descriptive statistics.
For each such quantity we compute its central tendency (the mean and the median), its extremes (the minimum and the maximum), and three measures of dispersion.
The dispersion measures are the standard deviation (SD), the interquartile range (IQR, the spread of the middle 50\% of values, which is robust to outliers), and the coefficient of variation (CV, the SD divided by the mean, a scale-free measure of relative variability).
For a few quantities we additionally compute the distribution shape: its skewness (asymmetry) and kurtosis (tailedness).
The scheme follows the measurement goal: the central-tendency statistics capture how fast, large, or forceful the writing is, the extremes capture its range, and the dispersion statistics capture its consistency.

The features fall into four families.

\emph{Kinematic} features summarize the speed and smoothness of movement.
Velocity, the instantaneous speed of the pen and hence the first time derivative of position, is computed in magnitude and along the horizontal ($x$) and vertical ($y$) axes and summarized by the statistics above; its central tendency indexes writing speed, and its dispersion indexes tempo consistency \citep{drotar2020dysgraphia, rosenblum2003computerized, smits1997dysgraphia}.
Acceleration and jerk, the first and second time derivatives of velocity, index movement control; jerk in particular is a standard smoothness measure that smooth, well-controlled movement minimizes \citep{flash1985coordination}.
We further compute the spectral arc length (SPARC), a dimensionless smoothness index obtained from the Fourier magnitude spectrum of the speed profile \citep{balasubramanian2015analysis}, and the number of velocity and acceleration changes (NCV and NCA), the counts of sign reversals in the velocity and acceleration signals, which index movement (dis)fluency \citep{amini2023identifications}.

\emph{Temporal} features summarize the time course of writing: the time the pen spends on the surface versus in the air (and their ratio), the durations of individual pen-down strokes, and the count, frequency, and durations of in-air pauses, which reflect hesitation and planning load \citep{rosenblum2003air, kunhoth2023exploration}.

\emph{Spatial} features summarize the size and geometry of the trace: overall writing extent (bounding-box width, bounding-box height, total path length); per-stroke length, height, width, and their regularity \citep{fallah2025ai, kunhoth2026multimodal, drotar2020dysgraphia}; and path geometry, namely straightness (net displacement relative to path length) together with the rate of direction changes \citep{okamoto1999line}.

\emph{Dynamic} features summarize pen forces and orientation: pen-tip pressure with its within-stroke modulation, which indexes fine force control \citep{drotar2016evaluation}; and pen tilt (altitude angle) and rotation (azimuth angle), which index how the pen is held \citep{asselborn2020extending}.

The pipeline is fully deterministic and has no learned parameters.

\subsubsection{Pre-registered 12-domain vocabulary}
\label{sec:vocab}
On their own, the 136 features form a flat vector that is hard to interpret: it mixes many correlated measurements without indicating which aspect of the writing process each one reflects.
To make the feature set interpretable and reusable, we organize it into a \emph{vocabulary} of 12 handwriting-process domains, each a named construct grounded in the motor-control and handwriting literature (Table~\ref{tab:vocab}).
Every feature is assigned to exactly one domain according to the process it indexes, so that the features within a domain all measure the same underlying construct.
The 12 domains are Writing Speed, Speed Consistency, Movement Smoothness, Temporal Organization, In-Air Behavior, Stroke Timing, Stroke Size, Writing Extent, Path Geometry, Pressure Control, Pen Tilt, and Pen Rotation.

The vocabulary is constructed from the literature alone, and the 12-domain $\times$ 136-feature mapping was fixed before any feature extraction or analysis of the cohort, so it cannot be retrofitted to the data \citep{nosek2018preregistration}.
This rests on three layers: (i) a deterministic feature pipeline with no learned parameters; (ii) a literature-only assignment of features to domains, anchored to named sources for each domain; and (iii) freezing of the mapping prior to feature extraction and analysis.

\begin{table*}[t]
\centering
\caption{Pre-registered 12-domain handwriting-process vocabulary: the 136
deterministic online-handwriting features organized into 12 motor-control domains
($k$ = features per domain). ``stats'' abbreviates the seven summary statistics
(mean, SD, median, max, min, IQR, CV) of the named quantity; feature definitions
and references are given in the text.}
\label{tab:vocab}
\small
\begin{tabularx}{\textwidth}{@{}llcX@{}}
\toprule
 & Domain & $k$ & Features \\
\midrule
D01 & Writing Speed & 14 & Velocity, globally and for the $x$/$y$ components (mean, median, max, min); overall writing speed; mean stroke velocity \\
D02 & Speed Consistency & 12 & Velocity variability, globally and for $x$/$y$ (SD, IQR, CV); velocity skewness, kurtosis, regularity \\
D03 & Movement Smoothness & 23 & Acceleration stats (plus skewness and kurtosis); jerk stats; number of velocity/acceleration changes (NCV, NCA); spectral arc length (velocity, $x$, $y$); normalized jerk; jerk per unit length \\
D04 & Temporal Organization & 4 & On-surface time and ratio; in-air time; air/surface ratio \\
D05 & In-Air Behavior & 17 & In-air stroke-length stats; total in-air duration; pause count, pause frequency, and pause-duration stats \\
D06 & Stroke Timing & 8 & Stroke-duration stats; stroke regularity \\
D07 & Stroke Size & 22 & Per-stroke length, height, and width stats; size regularity \\
D08 & Writing Extent & 5 & Overall writing width and height; total, horizontal, and vertical path distance \\
D09 & Path Geometry & 3 & Path straightness; mean stroke straightness; direction-change rate \\
D10 & Pressure Control & 14 & Pen-pressure stats (plus skewness and kurtosis); pressure--velocity correlation; pressure at stroke start and end; within-stroke pressure change; velocity--pressure ratio \\
D11 & Pen Tilt & 7 & Pen-altitude stats \\
D12 & Pen Rotation & 7 & Pen-azimuth stats \\
\midrule
\multicolumn{2}{@{}l}{Total} & 136 & \\
\botrule
\end{tabularx}
\end{table*}

\subsection{Reference-based deviation profiling}
We estimate expected performance from verified-typical (healthy) children only, and take each child's standardized residual from that model as their deviation \citep{van2013establishing, marquand2016understanding}.
Deriving the reference from non-pathological children alone follows the logic of pediatric growth charts \citep{cole1990lms}.
Aggregated within each domain, these deviations yield a per-child, 12-axis handwriting-process profile, the measurement the instrument produces.

For each feature $f$, we fit an ordinary least-squares (OLS) regression on the healthy children, using age and sex as predictors:
\begin{equation}
\label{eq:ols}
f_i = \beta_0 + \beta_1\,\text{age}_i + \beta_2\,\text{sex}_i + \varepsilon_i,
\quad i \in \mathrm{healthy}.
\end{equation}
We fit on healthy children only because including dysgraphic children would inflate the reference variance and shrink the very deviations we aim to measure.
We use a linear age term for parsimony given the reference size ($N=110$ healthy); it serves only to adjust for age and implies no claim that the age--feature relationship is exactly linear.

We then compute each child's residual against the healthy reference,
\begin{equation}
\label{eq:residual}
r_i^f = f_i - \big(\hat{\beta}_0 + \hat{\beta}_1\,\text{age}_i
        + \hat{\beta}_2\,\text{sex}_i\big),
\end{equation}
standardize it by the healthy residual standard deviation $\hat{\sigma}_f = \mathrm{std}(r_i^f : i \in \mathrm{healthy})$ to obtain a per-feature $z$-score $z_i^f = r_i^f / \hat{\sigma}_f$ (with a zero-variance guard, $z_i^f := 0$ when $\hat{\sigma}_f < 10^{-8}$), and aggregate to each domain $D$ by averaging the $z$-scores of its features:
\begin{equation}
\label{eq:meanz}
\overline{z}_i^{\,D} = \frac{1}{|D|} \sum_{f \in D} z_i^f .
\end{equation}
For each (child, domain) we additionally report a sign-coherence rate (SCR), the proportion of features in the domain that deviate in the same direction, as a descriptive measure of within-domain agreement.

Because the healthy reference is small, we report bootstrap 95\% confidence intervals ($B=1000$, resampling the healthy reference and refitting the full pipeline) to quantify reference-induced uncertainty.
When scoring healthy children themselves, we use a leave-one-subject-out (LOSO) refit, since scoring a child against a reference that includes that child would artificially shrink the residual.

\subsection{Validation analyses}
\label{sec:validation-design}
We establish the instrument through four validation analyses, one per measurement property (Aims~1--4).
All analyses are computed on the pooled cohort using the reference framework defined above.
Effect sizes are accompanied by bootstrap 95\% confidence intervals.
Where many hypotheses are tested at once, as in the known-groups contrasts across the twelve domains and their feature-level replication, we additionally apply Benjamini--Hochberg false-discovery-rate control.
Their results are reported in Section~\ref{sec:results}.

\subsubsection{Aim 1 -- Vocabulary establishment (E1)}
Building on the literature-grounded construction of the vocabulary (Section~\ref{sec:vocab}), we ask whether its 12 domains form a structurally coherent organization of the 136 features rather than an arbitrary grouping.

The primary indicator is \emph{intra-domain coherence}.
For each domain, we compute the Pearson correlation of every within-domain feature pair on the healthy children only.
We then summarize the domain by the mean of these correlations and report their full distribution.
A bootstrap 95\% confidence interval (resampling the healthy reference) quantifies its precision, and we regard a domain as coherent when this interval lies above zero.
This is a deliberately minimal criterion: the domains are formative groupings of complementary features rather than reflective scales whose items must intercorrelate highly, so the relevant question is whether within-domain features covary reliably at all \citep{bollen1991conventional}.
We restrict coherence estimation to healthy scores so that the structure reflects typical co-variation rather than between-group differences.

We then summarize the global structure as a $12 \times 12$ \emph{cross-domain similarity matrix}, whose diagonal entries are the intra-domain coherences above and whose off-diagonal entries are the correlations between the domains' mean deviation ($\overline{z}$) scores.
We expect a strong diagonal and weak but non-zero off-diagonal entries, because neighboring motor functions such as writing speed and stroke timing share variance.
The gap between the diagonal and off-diagonal magnitudes indexes how distinct the domains are.

\subsubsection{Aim 2 -- Reference framework validation (E2)}
We next examine whether the age- and sex-adjusted, healthy-only reference is well-calibrated, leakage-free under inferential testing, and parsimonious for the available sample size.

\emph{Healthy-on-healthy calibration.}
We apply the LOSO-refit reference to every healthy child and check that the resulting per-domain deviation scores behave as their construction implies.
Each feature $z$-score is a healthy residual divided by the healthy residual standard deviation, so on healthy children it is centered on zero with unit variance.
A domain score averages $k$ such correlated scores, so it remains centered on zero but is less dispersed, with an expected standard deviation of $\sqrt{(1+(k-1)\bar{r})/k}$ for a mean inter-feature correlation $\bar{r}$.
We therefore check four quantities: systematic bias (the per-domain mean offset), scaling (the observed healthy standard deviation against that analytical expectation), normality of the healthy domain scores (Shapiro--Wilk), which governs whether a $z$ value on this scale can be read as a normal-curve quantile, and the rate of extreme deviations among healthy children ($|\overline{z}|>1.96$).

\emph{Leakage.}
Scoring a healthy child against a reference that already includes that child shrinks the child's own residual and deflates their apparent deviation.
To quantify this self-inclusion bias we score every healthy child twice.
The in-sample score uses a reference fit on all healthy children.
The LOSO score, used throughout the paper, uses a reference fit on the remaining healthy children.
For each domain we then compare the mean absolute domain score under the two references.

\emph{Parsimony.}
We justify the linear age term, per feature, against quadratic, natural-spline, and age$\times$sex-interaction,
\begin{align}
\text{quadratic:}\quad & \beta_0 + \beta_1\,\text{age}_i + \beta_2\,\text{age}_i^2 + \beta_3\,\text{sex}_i, \nonumber\\
\text{spline:}\quad & \beta_0 + \sum_{m=1}^{4}\beta_m\,s_m(\text{age}_i) + \beta_5\,\text{sex}_i, \nonumber\\
\text{interaction:}\quad & \beta_0 + \beta_1\,\text{age}_i + \beta_2\,\text{sex}_i + \beta_3\,(\text{age}_i\cdot\text{sex}_i), \nonumber
\end{align}
where $\{s_m(\cdot)\}_{m=1}^{4}$ is a restricted (natural) cubic spline basis in age with five knots placed at the quantiles recommended by \citet{harrell2015regression}.
A restricted cubic spline is piecewise cubic between knots but is constrained to be linear beyond the outermost pair, which keeps the fit from swinging in the sparse tails of the age distribution; five knots give the four basis functions above.
Two complementary criteria are used.
The Bayesian Information Criterion (BIC) \citep{schwarz1978estimating} scores a model by its in-sample fit penalized for the number of parameters ($\mathrm{BIC}=-2\ln\hat{L}+p\ln n$, for a model with $p$ parameters fit on $n$ healthy children with maximized likelihood $\hat{L}$), so a more flexible model is preferred only when its improved fit outweighs its added complexity.
Per feature we select the model with the lowest BIC.
As an out-of-sample guard against overfitting, we compute the leave-one-out cross-validated mean squared error (LOOCV MSE) \citep{stone1974cross}: for each healthy child we refit the model on the remaining children, predict the held-out feature value, and average the squared prediction errors over all children, with smaller values indicating better generalization.

\subsubsection{Aim 3 -- Cross-sectional clinical validity (E3)}
We then test whether the domain deviation scores carry clinically meaningful signal, through known-groups validity.

\emph{Known-groups validity.}
For each domain we compare the dysgraphic and healthy deviation ($\overline{z}$) distributions with Cliff's $\delta$ \citep{cliff1993dominance}, a non-parametric effect size measuring how often a dysgraphic score exceeds a healthy score, and test significance with the Mann-Whitney $U$ test.
Because 12 domains are tested simultaneously, we control the false-discovery rate with the Benjamini--Hochberg procedure \citep{benjamini1995controlling} and report corrected $q$-values.

\emph{Stability of the effect-size ordering.}
To confirm these effect sizes are not artifacts of which healthy children constitute the reference, we bootstrap-resample the healthy reference ($B=1000$); in each sample we recompute the twelve per-domain $\delta$ values, re-rank the domains by $|\delta|$, and correlate each resample's ranking with the point-estimate ranking (Spearman).

\emph{Robustness to the protocol--age confound.}
Because protocol is collinear with age, the pooled dysgraphic--healthy contrast could in principle reflect protocol or age rather than group, so we subject it to a sensitivity suite.
We first quantify the marginal protocol$\times$group imbalance (Fisher's exact test and Cohen's $h$ \citep{cohen2013statistical}) and the protocol--age collinearity.
The primary test replicates the known-groups contrasts within each protocol and assesses their concordance with the pooled estimates through the agreement in sign and the rank correlation of the per-domain effects.
We further repeat the contrasts on the age-overlapping subsample and add a protocol indicator as a covariate to bound its residual effect.

\subsubsection{Aim 4 -- Individual reliability and interpretation (E4)}

We finally turn to the individual level.
We first ask whether a single diagnostic label conceals multidimensional profile heterogeneity that no single aggregate score can represent, and then whether the individual deviation profiles are reliable and interpretable enough to support per-child feedback and intervention-planning hypotheses.

\emph{Within-group heterogeneity.}
We first quantify the overall spread by comparing the mean pairwise Euclidean distance among dysgraphic profiles with the mean dysgraphic--healthy distance; a large within-group spread indicates that a single group label conceals distinct individual profiles.
A Euclidean distance, however, conflates a child's overall deviation magnitude (the profile's Euclidean length, a descriptive quantity and not a validated clinical severity rating) with its shape (which domains deviate).
We therefore decompose the two, to test whether the heterogeneity is multidimensional rather than a one-dimensional spread in magnitude.
Dividing out magnitude, we represent shape as the unit-length profile direction and compare the shape similarity (cosine) of magnitude-matched child pairs against that of all pairs; we also report the within-group inter-domain correlations and the per-child most-deviating domain (Appendix~\ref{app:shape}).
Both analyses are descriptive, reported without inferential testing, and motivate the individual-level profiling that follows.

\emph{Per-child reliability.}
For each child we compute a bootstrap 95\% confidence interval on all 12 domain scores by resampling the healthy reference and recomputing the full LOSO-refit pipeline.
We then summarize the distribution of interval widths across children and domains, for example the proportion of profiles narrower than one z-unit, and relate width to domain size and intra-domain coherence.
A narrow interval means a child's deviation is stable across plausible reference samples, so individualized feedback on that domain is well supported; a wide interval flags a score whose magnitude is uncertain.

\emph{Profile gallery.}
We present a gallery of eight individual profiles, selected by a deterministic rule keyed to an external criterion.
We stratify dysgraphic children by HPSQ-C legibility quartile, take the within-quartile median child with ties broken by ascending identifier, and match each to a healthy child of the same sex within 0.5 years of age.
The legibility quartile is deliberately not a motor-severity ranking.
Each profile is read through a per-domain, hypothesis-generating interpretation guide (Appendix~\ref{app:interp}).

\subsection{Reference implementation: system architecture and open API}
\label{sec:system}
\begin{sidewaysfigure}
\centering
\includegraphics[width=\textheight]{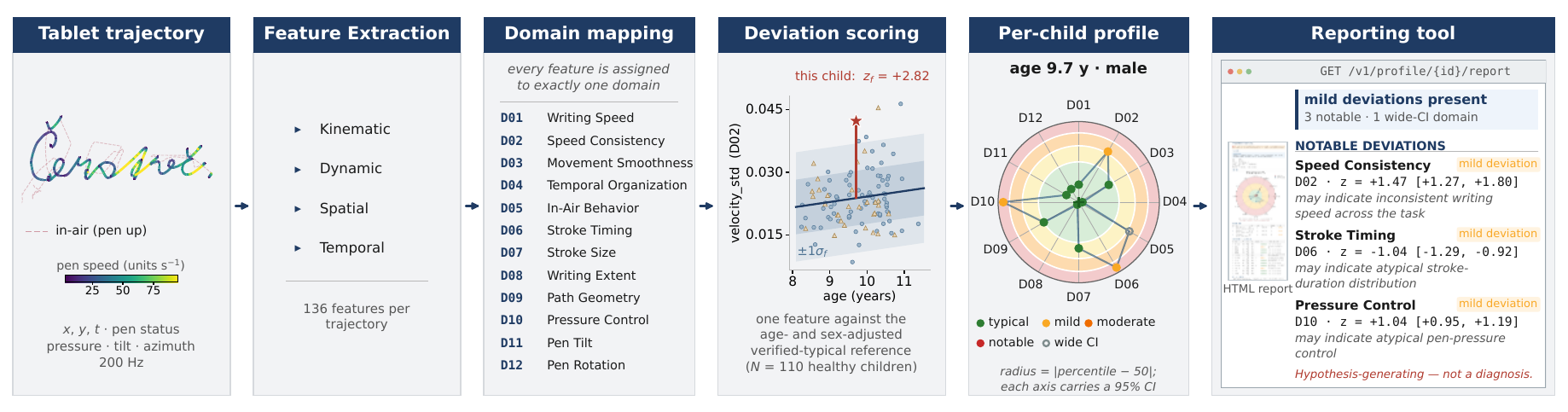}
\caption{The profiling pipeline and its open reference implementation, shown
end to end on one participant (9.7 years; male; DiaGraMo).
\emph{Tablet trajectory}: the recorded pen trajectory, colored by
instantaneous pen speed, with in-air movement dashed; the first word of a
dictated sentence is shown.
\emph{Feature extraction}: a deterministic, versioned extractor returns 136
kinematic, dynamic, spatial and temporal features.
\emph{Domain mapping}: the frozen vocabulary assigns every feature to
exactly one of the twelve interpretable domains.
\emph{Deviation scoring}: every feature is referenced to the verified-typical
cohort ($N=110$; ages 8.2--11.4) by an age- and sex-adjusted linear model and
the $z$ scores are averaged within a domain; the scatter shows one D02 feature
with its fitted age trend, $\pm1\sigma_f$ and $\pm2\sigma_f$ bands, and this
child's residual (circles, healthy boys; triangles, healthy girls).
\emph{Per-child profile}: the response is a twelve-axis deviation profile in
which the radius is the distance of the domain percentile from the healthy
median, and every axis carries a bootstrap 95\% confidence interval, a
sign-coherence ratio and a reliability flag (hollow marker, wide interval).
\emph{Reporting tool}: the shipped HTML report at left, its verdict and
notable-deviation rows enlarged at right. Every box is produced by the
reference implementation from the fixtures released with it.}
\label{fig:system}
\end{sidewaysfigure}

To make the validated method reusable, we provide an open reference implementation that operationalizes the full pipeline.
Given a new participant's pen trajectory together with age and sex, it returns the per-child 12-axis deviation profile with bootstrap confidence intervals, per-domain sign-coherence descriptors, reliability flags, and literature-anchored interpretation hints (Figure~\ref{fig:system}).
The implementation is deterministic and versioned.
It fits nothing at request time, drawing instead on the Czech verified-typical reference artifact that it ships, and every response carries the schema version and a hash of the vocabulary, so a result can be traced to the exact mapping and reference that produced it.
It is openly available at \url{https://github.com/jihyunmun/handwriting-process-profiling} and archived on Zenodo (\url{https://doi.org/10.5281/zenodo.22746861}).
As a supporting practical-use layer, it lets other researchers compute the same measure on new data; its illustrative use is shown in Section~\ref{sec:refimpl-use}.

%% file: sections/3_results.tex

\section{Results}
\label{sec:results}
We organize the results around the four measurement properties that establish the instrument (Aims~1--4): structural coherence of the 12-domain vocabulary (Aim~1); calibration and parsimony of the age- and sex-adjusted reference, with leakage-free inference (Aim~2); cross-sectional construct validity via known-groups separation (Aim~3); and individual-profile reliability with hypothesis-generating interpretation (Aim~4).
We then illustrate the open reference implementation on representative children, as a supporting practical-use layer rather than a further validation aim.

\subsection{Aim 1 -- Vocabulary establishment}
\label{sec:res-e1}
Intra-domain coherence and cross-domain similarity together indicate that the 12-domain vocabulary is a largely coherent structural organization of the 136 features, not an arbitrary grouping.

\subsubsection{Intra-domain coherence}
On healthy children, ten of the twelve domains have a confidence interval strictly above zero (Table~\ref{tab:e1a}), so their features covary as a group.
Coherence is strongest for D08~Writing~Extent ($r=+0.587$) and D12~Pen~Rotation ($+0.415$).
The other coherent domains show weaker positive correlations, down to $+0.045$ for D10~Pressure~Control, so the intervals establish reliably positive but often modest within-domain covariation rather than high internal consistency.
Two domains are exceptions.
D04~Temporal~Organization shows a small negative mean correlation ($-0.064$); this is a structural identity artifact rather than a coherence failure, because its on-surface and in-air time proportions are constrained to sum to one and are therefore negatively correlated by construction.
D09~Path~Geometry has a confidence interval that crosses zero ($+0.057$~[$-0.024$,~$+0.142$]), so its internal cohesion is not established.

\begin{table}[t]
\centering
\caption{Aim~1 intra-domain structural coherence: mean pairwise Pearson
correlation among each domain's features on healthy children ($N=110$), with
bootstrap 95\% CIs ($B=1000$). $k$ = number of features. Ten of twelve
domains have a CI strictly above zero.}
\label{tab:e1a}
\small
\begin{tabular}{@{}llcc@{}}
\toprule
Domain & $k$ & mean $r$ & 95\% CI \\
\midrule
D08 Writing Extent      & 5  & $+0.587$ & [$+0.503$, $+0.659$] \\
D12 Pen Rotation        & 7  & $+0.415$ & [$+0.337$, $+0.494$] \\
D07 Stroke Size         & 22 & $+0.232$ & [$+0.190$, $+0.268$] \\
D11 Pen Tilt            & 7  & $+0.220$ & [$+0.090$, $+0.368$] \\
D02 Speed Consistency   & 12 & $+0.210$ & [$+0.159$, $+0.254$] \\
D06 Stroke Timing       & 8  & $+0.186$ & [$+0.156$, $+0.216$] \\
D01 Writing Speed       & 14 & $+0.146$ & [$+0.104$, $+0.184$] \\
D05 In-Air Behavior     & 17 & $+0.118$ & [$+0.086$, $+0.158$] \\
D03 Movement Smoothness & 23 & $+0.109$ & [$+0.073$, $+0.144$] \\
D10 Pressure Control    & 14 & $+0.045$ & [$+0.024$, $+0.071$] \\
D09 Path Geometry       & 3  & $+0.057$ & [$-0.024$, $+0.142$]$^{\dagger}$ \\
D04 Temporal Org.       & 4  & $-0.064$ & [$-0.095$, $-0.040$]$^{\ddagger}$ \\
\botrule
\end{tabular}
\begin{flushleft}\footnotesize
$^{\dagger}$CI crosses zero (only three features; cohesion not claimed).
$^{\ddagger}$Structural identity artifact (on-surface + in-air proportions sum
to one); not a coherence failure.
\end{flushleft}
\end{table}

\subsubsection{Cross-domain similarity}
\begin{figure*}[!t]
\centering
\includegraphics[width=\textwidth]{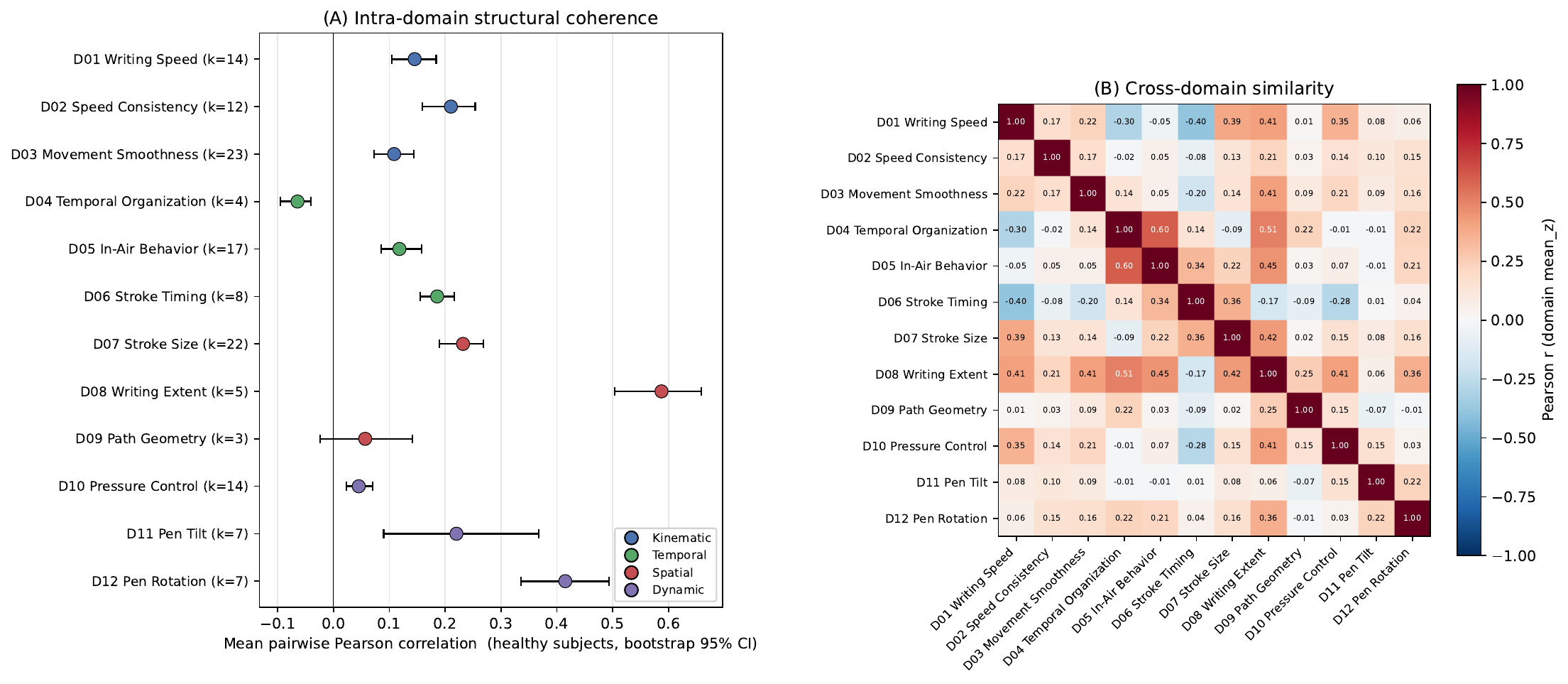}
\caption{Pre-registered 12-domain handwriting-process vocabulary:
(a) intra-domain structural coherence colored by family,
(b) cross-domain similarity matrix.}
\label{fig:vocab}
\end{figure*}

The 12$\times$12 cross-domain similarity matrix (Fig.~\ref{fig:vocab}b) confirms that the domains are distinct but partially related rather than orthogonal.
The largest off-diagonal similarity is D04~Temporal~Organization~$\leftrightarrow$~D05~In-Air~Behavior ($r\approx+0.60$), which share no features: D04 measures how on-surface and in-air time are apportioned, D05 what the pen does while off the surface.
Writing Extent (D08) shows moderate correlations with several domains (D01, D03--D05, D07, D10, D12; $r=0.36$--$0.51$).
All remaining off-diagonal correlations are moderate.

\subsection{Aim 2 -- Reference framework validation}
\label{sec:res-e2}
The age- and sex-adjusted, healthy-only reference is well-calibrated, leakage-free under inferential testing, and parsimonious at the available sample size.

\subsubsection{Healthy-on-healthy calibration}
Scoring healthy children against their own reference under leave-one-subject-out (LOSO) refitting yields well-calibrated domain scores (Table~\ref{tab:e2a}; Fig.~\ref{fig:s-calib}).
No domain mean departs from zero by more than $0.016$, so the reference introduces no systematic bias.
The observed standard deviations are all below one and follow the predicted pattern, largest for the few-feature, high-coherence domains (D08) and smallest for the many-feature, weakly correlated ones (D10).
Their median absolute departure from the prediction is $6.8\%$, about the standard error of a standard deviation estimated on 110 children; the two largest, D03~Movement~Smoothness ($+22\%$) and D05~In-Air~Behavior ($+14\%$), are the domains whose observed dispersion most exceeds what their coherence implies.

Extreme scores are rare among healthy children.
Per domain, at most two of the 110 healthy children exceed $|\overline{z}|>1.96$ (a rate of $1.8\%$; Table~\ref{tab:e2a}), and eight children ($7.3\%$) exceed it on at least one of the twelve domains.
These counts are descriptive: the domain scores have standard deviations well below one and, apart from D10~Pressure~Control, depart from normality (Shapiro--Wilk $p<.05$; Table~\ref{tab:e2a}), so the nominal $5\%$ attached to $|\overline{z}|>1.96$ does not apply.
Elsewhere the instrument locates a child by empirical percentiles of the verified-typical distribution rather than by this threshold (\S\ref{sec:individual-limits}).

\begin{table}[t]
\centering
\caption{Aim~2 healthy-on-healthy calibration under LOSO. Every domain
mean is within $\pm0.016$ of zero (no systematic bias); the observed
healthy standard deviation is compared with $\sqrt{(1+(k-1)\rho)/k}$,
evaluated at the intra-domain coherence $\rho$ of Table~\ref{tab:e1a}; and the
extreme-score rate ($|\overline{z}|>1.96$ among the 110 healthy children) is
low in every domain. Shapiro--Wilk $p$ tests the normality of the healthy
domain scores; only D10 is consistent with normality, so a $z$ value on this
scale is not read as a normal-curve quantile.
$^{\dagger}$D04's coherence is negative by construction,
so $\rho$ is set to zero for its prediction.}
\label{tab:e2a}
\footnotesize
\begin{tabular}{@{}lcccccc@{}}
\toprule
Domain & $k$ & Mean $\overline{z}$ & Exp.\ std & Obs.\ std & Outlier \% & S--W $p$ \\
\midrule
D01 Writing Speed       & 14 & $-0.004$ & 0.454 & 0.437 & 0.0 & $.020$ \\
D02 Speed Consistency   & 12 & $+0.008$ & 0.526 & 0.560 & 0.9 & $<.001$ \\
D03 Movement Smoothness & 23 & $+0.016$ & 0.384 & 0.467 & 0.9 & $<.001$ \\
D04 Temporal Org.       & 4  & $+0.005$ & 0.500$^{\dagger}$ & 0.475 & 0.9 & $<.001$ \\
D05 In-Air Behavior     & 17 & $+0.014$ & 0.412 & 0.471 & 0.9 & $<.001$ \\
D06 Stroke Timing       & 8  & $+0.009$ & 0.536 & 0.586 & 0.9 & $<.001$ \\
D07 Stroke Size         & 22 & $+0.013$ & 0.517 & 0.571 & 0.9 & $<.001$ \\
D08 Writing Extent      & 5  & $+0.005$ & 0.819 & 0.877 & 1.8 & $<.001$ \\
D09 Path Geometry       & 3  & $+0.010$ & 0.609 & 0.658 & 1.8 & $<.001$ \\
D10 Pressure Control    & 14 & $+0.002$ & 0.337 & 0.336 & 0.0 & $.244$ \\
D11 Pen Tilt            & 7  & $+0.010$ & 0.576 & 0.549 & 0.9 & $<.001$ \\
D12 Pen Rotation        & 7  & $+0.010$ & 0.706 & 0.669 & 0.9 & $<.001$ \\
\botrule
\end{tabular}
\end{table}

\subsubsection{Leakage-free inference}
Scoring under the LOSO refit removes self-inclusion by construction; relative to an in-sample reference, it enlarges the mean healthy $|\overline{z}|$ by a median of $0.022$ and at most $0.026$ (D08~Writing~Extent) across the twelve domains (per-domain detail in Appendix~\ref{app:refdetail}, Table~\ref{tab:s-leak}).

\subsubsection{Parsimony of the age model}
The linear age model is the appropriate parsimony choice at this reference size (Table~\ref{tab:e2b}).
Across the 129 features with non-zero variance, a linear \texttt{age + sex} reference is preferred by BIC over quadratic, restricted-cubic-spline, and age$\times$sex-interaction alternatives for 118 features (91.5\%), and by closed-form leave-one-out cross-validation for 66 features (51.2\%).
For all three flexible alternatives, the median change in LOOCV error relative to the linear model is positive, and so worse out-of-sample (quadratic $+0.013$, spline $+0.025$, interaction $+0.011$).
The near-even LOOCV split and these negligible median differences indicate that the flexible models are predictively equivalent to the linear one, so the strong BIC preference for the simple specification is decisive.

\begin{table}[t]
\centering
\caption{Aim~2 parsimony of the age model. Per-feature model selection
on healthy children ($N=110$), 129 non-zero-variance features. A positive
median $\Delta$LOOCV means the flexible model predicts \emph{worse} out-of-sample
than the linear model, so the value favors the linear reference.}
\label{tab:e2b}
\small
\begin{tabular}{@{}lc@{}}
\toprule
Metric & Value \\
\midrule
Features BIC-preferring linear      & 118 / 129 (91.5\%) \\
Features LOOCV-preferring linear    & 66 / 129 (51.2\%) \\
Median $\Delta$LOOCV, quadratic vs.\ linear   & $+0.013$ (favors linear) \\
Median $\Delta$LOOCV, spline vs.\ linear      & $+0.025$ (favors linear) \\
Median $\Delta$LOOCV, age$\times$sex vs.\ linear & $+0.011$ (favors linear) \\
\botrule
\end{tabular}
\end{table}

\subsection{Aim 3 -- Cross-sectional clinical validity}
\label{sec:res-e3}
The domain deviation scores separate dysgraphic from typically-developing children (known-groups validity), and the separation is stable under resampling of the healthy reference and robust to the protocol--age confound.

\subsubsection{Known-groups validity}
\label{sec:res-e3a}
\begin{figure*}[t]
\centering
\includegraphics[width=\textwidth]{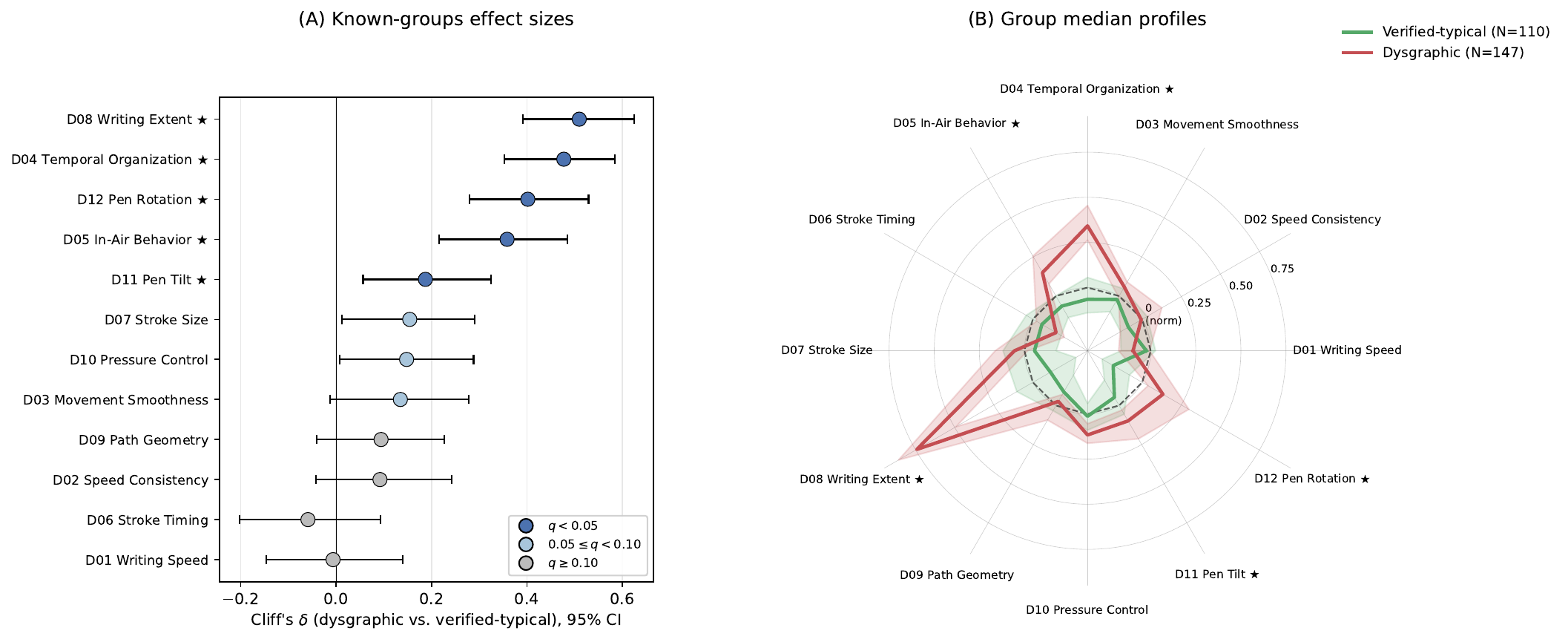}
\caption{Aim~3 known-groups validity. \textbf{(A)} Per-domain Cliff's $\delta$
(dysgraphic vs.\ verified-typical) with 95\% bootstrap CI ($B=1000$), domains
ordered by $|\delta|$; marker shade encodes Benjamini--Hochberg significance over
the 12 domains (dark: $q<0.05$; light: $0.05\le q<0.10$; gray: $q\ge0.10$), and
$\bigstar$ marks the five significant domains. All significant effects are
positive, indicating deviation above the age- and sex-adjusted healthy reference.
\textbf{(B)} The same signal as group profiles: the median per-domain deviation
($\overline{z}$) of the dysgraphic group (red) and the verified-typical group
(green), each with a bootstrap 95\% CI band, in the 12-domain space (dashed ring =
verified-typical norm). The dysgraphic median profile departs from the near-zero
typical profile on exactly those five axes and overlaps it elsewhere. Panel~B uses
medians because they are robust to the heavy right tail of D02~Speed~Consistency
(the least-reliable domain), whose group \emph{mean} is inflated by two outliers
but whose median, like its known-groups $\delta$, is null.}
\label{fig:e3a}
\end{figure*}

Deviation scores separate the dysgraphic and healthy groups on a subset of the domains (Table~\ref{tab:e3a}, Fig.~\ref{fig:e3a}).
Five of the twelve domains reach Benjamini--Hochberg significance at $q<0.05$, all with positive Cliff's $\delta$: D08~Writing~Extent ($\delta=+0.509$), D04~Temporal~Organization ($+0.477$), D12~Pen~Rotation ($+0.402$), D05~In-Air~Behavior ($+0.358$), and D11~Pen~Tilt ($+0.187$).
Every significant effect is positive: dysgraphic children deviate \emph{above} the age- and sex-adjusted healthy reference on each of these domains.
Three further domains are near-significant ($0.05\le q<0.10$: D07~Stroke~Size, D10~Pressure~Control, D03~Movement~Smoothness) but do not survive correction.
The remaining four domains do not separate the groups: D09~Path~Geometry, D02~Speed~Consistency, D06~Stroke~Timing, and D01~Writing~Speed, which is essentially null ($\delta=-0.006$, $q=0.930$).
Rendered as group profiles against the verified-typical reference (Fig.~\ref{fig:e3a}B), these same five domains are precisely the axes on which the dysgraphic group's median deviation profile departs from the near-zero typical profile, while the remaining domains overlap it.
Repeating the contrast on individual features, without any averaging, shows that aggregation does not manufacture the domain-level signal.
Of the 136 features, 58 are significant under the same correction, and where they fall tracks the domain result: each of the five domain-significant domains has at least $57\%$ of its features individually significant, and D04 and D08 have all of them, against at most $21\%$ in the four null domains.
Aggregation does not manufacture the signal, though it can blunt a locally strong one: the single strongest feature, \texttt{jerk\_cv} ($\delta=-0.612$), lies in D03~Movement~Smoothness, a domain that only reaches near-significance once its features are averaged (Appendix~\ref{app:feature}, Table~\ref{tab:s-feature}).
Because dysgraphia frequently co-occurs with dyslexia, we further checked that the group signal is not driven by that comorbidity.
The dysgraphia-only subgroup ($N=115$) reproduces the pooled pattern, so the effect is not an artifact of co-occurring dyslexia, while the smaller dysgraphia-plus-dyslexia subgroup ($N=29$) shows amplified deviation on the temporal and path-geometry domains (D04 and D09), an exploratory observation at that subgroup size (Appendix~\ref{app:comorbid}, Table~\ref{tab:s-comorbid}).

\begin{table}[t]
\centering
\caption{Aim~3 known-groups validity. Per-domain Cliff's $\delta$
(dysgraphic vs.\ healthy) with bootstrap 95\% CI ($B=1000$), Mann--Whitney $U$
test, Benjamini--Hochberg-adjusted $q$ over 12 domains. Sorted by $|\delta|$.
$\checkmark$ = $q<0.05$; $\sim$ = near-significant ($0.05\le q<0.10$).}
\label{tab:e3a}
\small
\begin{tabular}{@{}lcccc@{}}
\toprule
Domain & Cliff's $\delta$ & 95\% CI & BH $q$ & Sig. \\
\midrule
D08 Writing Extent      & $+0.509$ & [$+0.391$, $+0.624$] & $3.4\mathrm{e}{-}11$ & $\checkmark$ \\
D04 Temporal Org.       & $+0.477$ & [$+0.353$, $+0.584$] & $3.7\mathrm{e}{-}10$ & $\checkmark$ \\
D12 Pen Rotation        & $+0.402$ & [$+0.279$, $+0.529$] & $1.5\mathrm{e}{-}07$ & $\checkmark$ \\
D05 In-Air Behavior     & $+0.358$ & [$+0.216$, $+0.485$] & $2.7\mathrm{e}{-}06$ & $\checkmark$ \\
D11 Pen Tilt            & $+0.187$ & [$+0.057$, $+0.325$] & $0.025$ & $\checkmark$ \\
D07 Stroke Size         & $+0.154$ & [$+0.012$, $+0.290$] & $0.069$ & $\sim$ \\
D10 Pressure Control    & $+0.148$ & [$+0.007$, $+0.288$] & $0.074$ & $\sim$ \\
D03 Movement Smoothness & $+0.135$ & [$-0.012$, $+0.279$] & $0.097$ & $\sim$ \\
D09 Path Geometry       & $+0.094$ & [$-0.041$, $+0.227$] & $0.250$ & \\
D02 Speed Consistency   & $+0.092$ & [$-0.042$, $+0.242$] & $0.250$ & \\
D06 Stroke Timing       & $-0.059$ & [$-0.203$, $+0.093$] & $0.457$ & \\
D01 Writing Speed       & $-0.006$ & [$-0.147$, $+0.140$] & $0.930$ & \\
\botrule
\end{tabular}
\end{table}

\subsubsection{Stability of the effect-size ordering}
The known-groups effect sizes are stable under resampling of the healthy reference: across $B=1000$ bootstrap resamples, the Spearman correlation between each resample's ordering of the twelve domains by $|\delta|$ and the point-estimate ordering has a median of $0.979$ (IQR $0.028$), so which domains carry the strongest group signal is essentially invariant to which healthy children constitute the reference (per-domain bootstrap CIs on $\delta$ in Appendix~\ref{app:refdetail}, Table~\ref{tab:s-stab}).
Whereas that check varies which healthy children form the reference, the per-domain $\delta$ is equally insensitive to how the reference is fitted.
Across in-sample versus LOSO scoring and a sex-stratified fit, in which the reference is fitted separately by sex instead of taking sex as a covariate, no domain's $\delta$ changes by more than $0.016$.
A further protocol-indicator variant is tabulated alongside these (Appendix~\ref{app:refdetail}, Table~\ref{tab:s-robust}) and examined directly in the confound analysis below (Section~\ref{sec:res-s2}).

\subsubsection{Robustness to the protocol--age confound}
\label{sec:res-s2}
The marginal protocol$\times$group imbalance is small and non-significant (Fisher exact $p=0.20$, Cohen's $h=0.17$), and the known-groups contrast holds up under the sensitivity suite for the protocol--age confound (Table~\ref{tab:s2}; Fig.~\ref{fig:s-protocol}).
The primary robustness test replicates the group contrasts \emph{within} each age-stratified protocol, where healthy and dysgraphic children are age-matched and perform identical tasks.
There, the sign of Cliff's $\delta$ agrees with the pooled analysis for 9 of 12 domains, and the rank correlation of $|\delta|$ between protocols is $0.769$.
The three sign disagreements (D01, D06, D09) all occur on small or null pooled effects ($|\delta|\le0.094$), i.e.\ sign flips in noise around zero; every domain with a non-negligible pooled effect agrees in direction across protocols.
Two corroborating analyses confirm this verdict.
Restricting to the age-overlap band where both protocols coexist (Fig.~\ref{fig:s-overlap}) changes each domain's $\delta$ by at most $0.109$ (median $0.045$) while preserving the sign of every effect.
The largest shift falls on the strongest domain, D08~Writing~Extent, whose effect nonetheless remains large ($\delta=0.51\rightarrow0.40$), and the next-largest on the null domain D09~Path~Geometry ($0.09\rightarrow0.00$).
Adding an explicit protocol indicator to the reference regression changes $\delta$ by at most $0.030$ (median $0.004$).
Because protocol assignment tracks age, the age term already accounts for most of what a protocol indicator could; what this check adds is that the remainder, chiefly the difference in tasks between the two protocols, does not materially move the group contrast either.
The five domains that reach significance hold their direction, and approximately their magnitude, under every analysis in the suite.
The three sign-unstable domains (D01, D06, D09) carry no established effect in either direction, and we draw no directional conclusion from them.

\begin{table}[t]
\centering
\caption{Aim~3 robustness: within-protocol replication of the
known-groups Cliff's $\delta$. Sign agreement $=9/12$;
rank correlation of $|\delta|$ between protocols $=0.769$. The three
disagreements (\textbf{N}) all fall on null/small pooled effects
($|\delta|\le0.094$).}
\label{tab:s2}
\small
\begin{tabular}{@{}lcccc@{}}
\toprule
Domain & Pooled $\delta$ & Protocol A & Protocol B & Sign \\
\midrule
D01 Writing Speed       & $-0.006$ & $-0.121$ & $+0.062$ & \textbf{N} \\
D02 Speed Consistency   & $+0.092$ & $+0.160$ & $+0.081$ & Y \\
D03 Movement Smoothness & $+0.135$ & $+0.291$ & $+0.100$ & Y \\
D04 Temporal Org.       & $+0.477$ & $+0.459$ & $+0.471$ & Y \\
D05 In-Air Behavior     & $+0.358$ & $+0.339$ & $+0.393$ & Y \\
D06 Stroke Timing       & $-0.059$ & $+0.004$ & $-0.043$ & \textbf{N} \\
D07 Stroke Size         & $+0.154$ & $+0.178$ & $+0.143$ & Y \\
D08 Writing Extent      & $+0.509$ & $+0.513$ & $+0.444$ & Y \\
D09 Path Geometry       & $+0.094$ & $+0.239$ & $-0.014$ & \textbf{N} \\
D10 Pressure Control    & $+0.148$ & $+0.222$ & $+0.087$ & Y \\
D11 Pen Tilt            & $+0.187$ & $+0.191$ & $+0.204$ & Y \\
D12 Pen Rotation        & $+0.402$ & $+0.326$ & $+0.432$ & Y \\
\botrule
\end{tabular}
\end{table}

\subsection{Aim 4 -- Individual reliability and interpretation}
\label{sec:res-e4}
Having established group-level validity, we turn to the individual level.
The premise of an individual profile is that a single binary diagnostic label discards structured information.
We first show that the within-label variation is large and multidimensional, so that neither the label nor any single aggregate score captures it, and then that the individual profiles are reliable and interpretable enough to support per-child use.

\subsubsection{Within-group profile heterogeneity}
\label{sec:res-e4het}
Pairwise Euclidean distances between children in the 12-dimensional profile space (Table~\ref{tab:e4het}) give a within-dysgraphic to cross-group distance ratio of $1.089$: two dysgraphic children are, on average, as different from each other as a dysgraphic child is from a healthy control.
A one-bit label cannot carry that structure.
The variation is also multidimensional rather than a spread in \emph{how much} children deviate overall.
Holding a child's overall deviation magnitude fixed, their profile \emph{shape} (which domains deviate) still varies almost as much as it does across all children.
D08~Writing~Extent is the most frequent peak domain, yet it is the largest deviation for only about a third of children; for the other $64\%$ the peak lies elsewhere (Appendix~\ref{app:shape}, Fig.~\ref{fig:s-shape}).
Two children can therefore share both the label and the same overall deviation yet be flagged on different handwriting processes, so neither a binary label nor a single aggregate deviation score can substitute for the profile.

\begin{table}[t]
\centering
\caption{Aim~4 within-group profile heterogeneity: pairwise Euclidean
distances in 12-dimensional profile space. The within-dysgraphic /
cross-group ratio is $1.089$.}
\label{tab:e4het}
\small
\begin{tabular}{@{}lccc@{}}
\toprule
Scope & \# pairs & Mean dist. & Median \\
\midrule
Within-dysgraphic & 10,731 & 3.498 & 2.707 \\
Within-healthy    & 5,995  & 2.583 & 2.426 \\
Cross-group       & 16,170 & 3.212 & 2.654 \\
\botrule
\end{tabular}
\end{table}

\subsubsection{Per-child reliability}
\label{sec:res-e4a}
Individual-level deviation scores are accompanied by bootstrap uncertainty from resampling the healthy reference (Table~\ref{tab:e4a}; Fig.~\ref{fig:s-ci}).
Across all child~$\times$~domain pairs, 96.5\% have a 95\% CI width below one $z$-score unit, so individual interpretation is supportable when the accompanying uncertainty is respected.
Because a domain score is an average of its feature $z$-scores, how precisely the reference pins it down depends on how many features it pools and how correlated they are.
Averaging \emph{many}, only \emph{weakly} correlated features cancels most of the reference sampling noise and gives a narrow interval; averaging a \emph{few}, \emph{strongly} correlated features cancels little of it and gives a wide one.
The two extremes bracket this pattern: D10~Pressure~Control, pooling 14 weakly correlated features, has the tightest intervals (mean width $0.223$; 100\% below $0.5$), whereas D08~Writing~Extent, pooling only 5 highly correlated features, has the widest (mean $0.694$; only 23\% below $0.5$).
D02~Speed~Consistency should not be read from its mean alone: the large \emph{mean} width ($0.789$) is inflated by a few children with very wide intervals, whereas its median ($0.376$) and its $79\%$ below $0.5$ place it among the more precise domains for most children.

\begin{table}[t]
\centering
\caption{Aim~4 per-child reliability: distribution of per-subject 95\%
bootstrap CI widths ($B=500$) across the 12 domains. Overall, 96.5\% of
child~$\times$~domain pairs have a CI width below 1.0. Domains are ordered by
mean width; D02~Speed~Consistency's large mean reflects a right tail of a few
wide-interval children, so its median ($0.376$) is the more representative summary.}
\label{tab:e4a}
\small
\begin{tabular}{@{}lcccc@{}}
\toprule
Domain & Mean & Median & \% $<0.5$ & \% $<1.0$ \\
\midrule
D10 Pressure Control    & 0.223 & 0.213 & 100 & 100 \\
D03 Movement Smoothness & 0.311 & 0.251 & 92  & 99 \\
D01 Writing Speed       & 0.374 & 0.306 & 89  & 97 \\
D04 Temporal Org.       & 0.383 & 0.322 & 82  & 98 \\
D05 In-Air Behavior     & 0.387 & 0.332 & 84  & 98 \\
D07 Stroke Size         & 0.399 & 0.350 & 89  & 98 \\
D06 Stroke Timing       & 0.426 & 0.373 & 80  & 98 \\
D11 Pen Tilt            & 0.456 & 0.427 & 70  & 98 \\
D09 Path Geometry       & 0.505 & 0.432 & 65  & 95 \\
D12 Pen Rotation        & 0.526 & 0.452 & 63  & 93 \\
D08 Writing Extent      & 0.694 & 0.609 & 23  & 88 \\
D02 Speed Consistency   & 0.789 & 0.376 & 79  & 96 \\
\botrule
\end{tabular}
\end{table}

\subsubsection{Pre-specified child profile gallery}
\label{sec:res-e4b}
\begin{figure*}[!t]
\centering
\includegraphics[width=\textwidth]{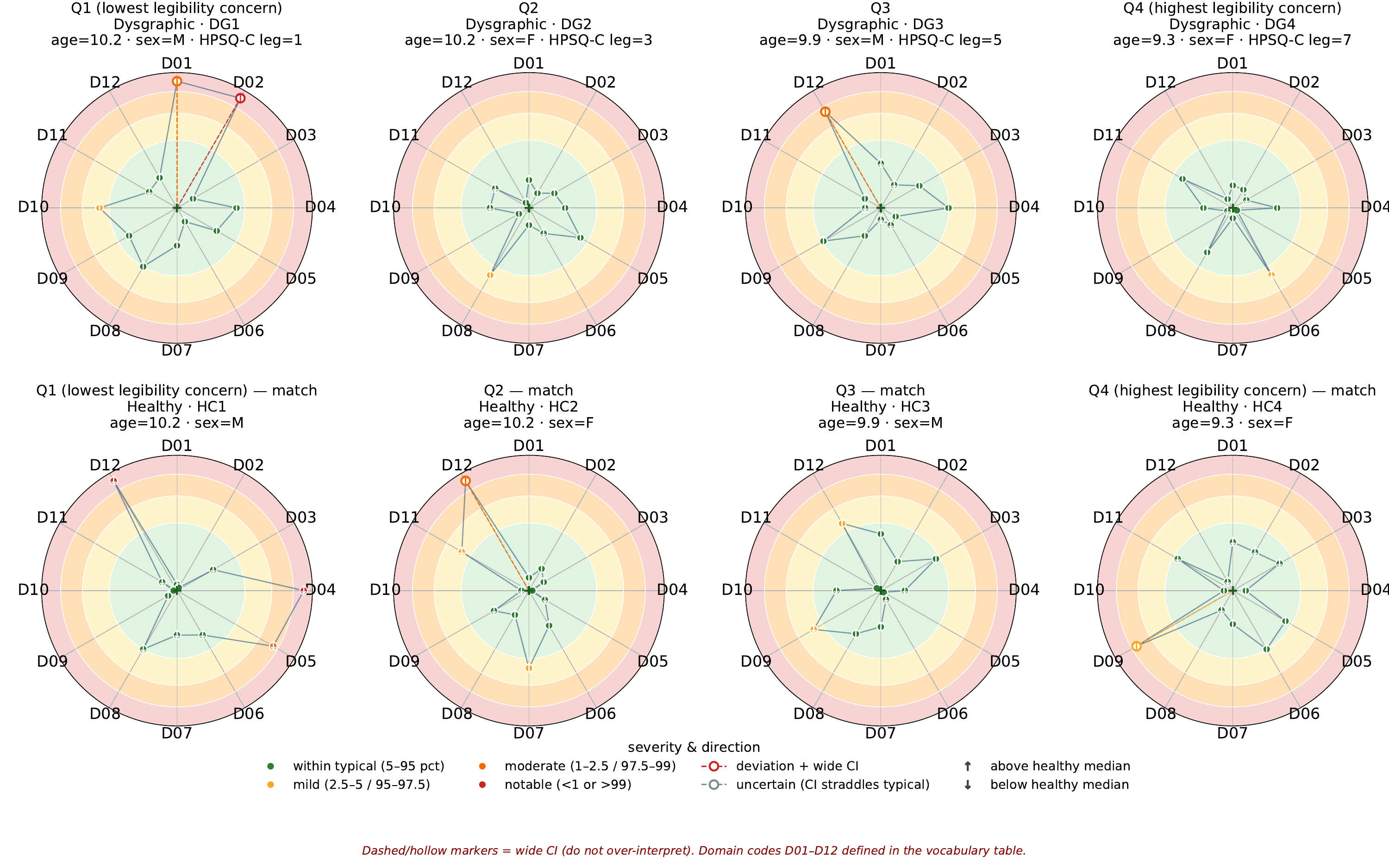}
\caption{Per-child 12-domain handwriting-process profiles (Aim 4), rendered in
the reference implementation's percentile--severity grammar. Top row: one
dysgraphic child per HPSQ-C legibility quartile (Q1--Q4), the within-quartile
median-legibility child. Bottom row: age- and sex-matched healthy controls. The
legibility quartile is an external parent-report concern axis used only to select
illustrative subjects reproducibly and non-circularly; it is \emph{not} a
motor-severity ordering. Radial coordinate: the domain's percentile in the
verified-typical distribution, drawn as $|\text{percentile}-50|$ (center =
typical; successive bands = mild / moderate / notable deviation); an
$\uparrow$/$\downarrow$ glyph marks above/below the healthy median; dashed, hollow
markers flag domains whose bootstrap CI is wide and should not be interpreted
individually. The dysgraphic children's largest apparent deviations (Q1 on
D02~Speed~Consistency and D01~Writing~Speed, Q3 on D12~Pen~Rotation) are all
wide-CI-flagged, and the matched healthy controls also show occasional
moderate-to-notable deviations (e.g.\ the Q1 control at the low tail of D04, D05,
and D12). Both are expected of a \emph{positional} instrument, not a diagnostic
classifier: some typical children necessarily fall in the tails of the
verified-typical distribution, and the interpretive weight is carried by deviation
direction and CI width rather than by magnitude alone (see \S\ref{sec:individual-limits}).}
\label{fig:gallery}
\end{figure*}

The eight profiles selected by the rule of Section~\ref{sec:validation-design} (subjects in Table~\ref{tab:e4b}) peak on different domains, and once the bootstrap intervals are respected none shows more than a mild deviation (Fig.~\ref{fig:gallery}).
The Q1 child (DG1) deviates chiefly on D02~Speed~Consistency and D01~Writing~Speed, and the Q3 child (DG3) on D12~Pen~Rotation, but each of these largest deviations carries a wide bootstrap CI and is flagged by the instrument as not individually interpretable.
The Q2 child (DG2) is typical except for a mild, reliably-estimated D08~Writing~Extent elevation, and the Q4 child (DG4) lies within the typical range on every domain but a mild D06~Stroke~Timing.
The matched healthy controls are not uniformly flat either: the Q1 control (HC1) sits at the low tail of D04~Temporal~Organization, D05~In-Air~Behavior, and D12~Pen~Rotation (all below the median), and the Q2 control (HC2) shows a wide-CI D12~Pen~Rotation elevation.
A typical child can therefore deviate on several domains just as a labeled child can.
Because the reference bands are percentiles of the verified-typical distribution, such tail deviations are an expected positional property; we take up their interpretation, and the limits of individual-level reading, in the Discussion (\S\ref{sec:individual-limits}).

\subsubsection{Hypothesis-generating interpretation guide}
\label{sec:res-e4c}
To make the profiles actionable while respecting the cross-sectional design, we provide a per-domain interpretation guide (Appendix Table~\ref{tab:s-interp}) that links each domain to its motor-control construct, a possible reading of a deviation (strict ``may indicate'' wording), and a literature-suggested direction for targeted assessment.
Every entry is hypothesis-generating: the guide never prescribes an intervention or makes a treatment-efficacy claim, and deviations are described as handwriting-process deviations rather than isolated motor deficits.

\subsection{Reference implementation: illustrative use}
\label{sec:refimpl-use}
As a supporting practical-use layer, and not a further validation aim, the open reference implementation renders the validated measure on new data.
We walk through two scenarios, both drawing on children shown in Fig.~\ref{fig:gallery} and rendered in the implementation's percentile--severity grammar, each using the same output to drive a different, strictly hypothesis-generating reading.

\subsubsection{U1 -- Initial profile triage}
For child DG1 (HPSQ-C legibility $=1$, Q1 in Fig.~\ref{fig:gallery}), the only domains outside the typical range are D02~Speed~Consistency and D01~Writing~Speed.
Both carry wide bootstrap CIs and are flagged as not individually interpretable, so no domain is reliably outside the typical range.
Despite the diagnostic label, then, the implementation surfaces no reliable per-domain target and defers to repeat measurement: a labeled child need not show a reliably-flagged profile signal, a distinction a binary label cannot make.

\subsubsection{U2 -- Intervention prioritization}
For child DG2 (HPSQ-C legibility $=3$, Q2 in Fig.~\ref{fig:gallery}), the profile is typical on eleven domains and flags a single reliably-estimated (narrow-CI) mild elevation on D08~Writing~Extent.
The implementation ranks domains by deviation and uncertainty and surfaces D08 as the one stable target; the accompanying guide suggests prioritizing assessment of spatial-extent regulation, and stops there rather than recommending an intervention.

%% file: sections/4_discussion.tex
\section{Discussion}

\subsection{From a single label to a validated process profile}
We set out to reframe the measurement of children's handwriting difficulty from a single categorical judgment to a graded, multidimensional profile.
Where prior approaches yield either a single global proficiency score or a binary dysgraphic-or-typical label, the instrument introduced here returns a twelve-axis, age- and sex-normed profile of handwriting-process deviations, each axis carrying a per-child bootstrap confidence interval and a strictly hypothesis-generating reading against a verified-typical reference.
Its unit of measurement is a position rather than a category: how far, and on which processes, a child's writing departs from that of typically-developing peers of the same age and sex.

Across four measurement properties, the validation evidence supports this profile as a usable measure.
The twelve-domain vocabulary is largely structurally coherent; the age- and sex-adjusted reference is well-calibrated, parsimonious, and leakage-free under inferential testing; the domain deviations carry known-groups construct validity, separating dysgraphic from typically-developing children on a subset of domains; and the individual profiles are reliable enough to support per-child interpretation, or, where their bootstrap uncertainty is too large, to decline it explicitly.
These claims rest on a measurement model that could not be fitted to the results it produced.
The vocabulary and its feature-to-domain mapping were fixed before any analysis of the cohort, behind three layers: a deterministic feature pipeline, a literature-only vocabulary, and a mapping frozen before feature extraction.

\subsection{What the four measurement properties establish}
Because the vocabulary and its feature-to-domain mapping were fixed from the literature before any cohort analysis, the observed intra-domain coherence corroborates a theory-driven grouping, not a data-driven one.
That the within-domain correlations are generally modest is expected.
The domains are formative groupings of complementary features chosen to capture a process from several angles, not reflective scales whose items must intercorrelate highly, so the appropriate structural indicator is reliably positive covariation rather than high internal consistency.
The two reported exceptions (\S\ref{sec:res-e1}) are definitional rather than disqualifying, and carry a methodological caution, since a pairwise-correlation criterion can misread domains built from compositional sets (D04, whose proportions sum to one) or feature-sparse ones (D09, with only three features).
What is provisional in both cases is the internal structure of the domain, not the group contrast it carries; the contrast is computed from the same domain score but does not require the features within a domain to cohere, which is why D04 can be among the strongest known-groups domains while its coherence statistic is uninformative.
Across domains, the vocabulary is also non-orthogonal by design: because the domains name interpretable, literature-grounded constructs and not statistically independent factors, neighboring functions and a global writing-size factor share variance without duplicating one another.
Interpretability is thus gained at the cost of strict independence, and the axes are safest read as a multi-domain pattern rather than as isolated scores.

The calibration, parsimony, and leakage analyses together establish the age- and sex-adjusted reference as well-behaved, with two consequences beyond the numbers themselves.
First, the sufficiency of a linear age term indicates near-linear age--feature relationships over this narrow age band, and the same term absorbs the protocol-linked variance that would otherwise confound the pooled comparison, which ties parsimony to the confound defense.
Second, the healthy children both define the reference and serve as the comparison group in the known-groups analysis, an arrangement that would be circular if a child's own data pulled their score toward zero.
Scoring healthy children under a leave-one-subject-out refit rules that out by construction, and the leakage analysis shows the effect would be slight in any case, since each child contributes about one observation in a hundred to the fit.
These properties hold for this cohort alone, because the calibration is specific to its healthy children and the sufficiency of a linear term is sample-size-dependent.

The construct-validity evidence is most informative in its pattern.
The domains that separate the groups are spatial, temporal-organizational, and pen-orientation processes: D08~Writing~Extent, D04~Temporal~Organization, D12~Pen~Rotation, D05~In-Air~Behavior, and D11~Pen~Tilt.
Every significant difference runs in the same direction, with dysgraphic children deviating further from the typical range rather than toward it.
Speed is absent from this list, which is itself informative, since speed is what the dominant assessment tradition foregrounds.
In this cohort, then, dysgraphia is distinguished not by how fast children write but by how they write: the size and spatial regulation of the trace, the allocation of on-surface and in-air time, and the orientation of the pen.

The non-discriminating domains refine this reading.
The speed null is not confined to one measure, since D02~Speed~Consistency is null as well as D01, so it is the speed construct as a whole that fails to separate the groups.
D06~Stroke~Timing is null while D04~Temporal~Organization is among the strongest domains, which locates the temporal signal in how a child apportions on-surface and in-air time rather than in the regularity of individual stroke durations.
D09~Path~Geometry's null is the least interpretable, since it is also the domain whose internal coherence could not be established, so its flat contrast may reflect a weak construct as much as an absent effect.

Aggregating features into interpretable domains costs some sensitivity.
The feature-level replication recovers the same processes, yet the single most discriminating feature lies in D03~Movement~Smoothness, a domain that is only near-significant once its features are averaged.
Its individually significant features split in direction, six positive and five negative, so averaging cancels signal that is clear feature by feature: dysgraphic children show higher mean jerk but lower jerk variability, and the domain score records neither.
That pattern is nonetheless robust.
It holds under resampling of the healthy reference, alternative reference-model specifications, and the protocol--age confound, and is therefore a property of the measure rather than of a particular reference sample or the age-stratified protocol split.
The near-significant domains are directionally consistent with the significant ones, but their effects are small and we read them as suggestive rather than established.

For most child--domain pairs the per-child bootstrap interval is narrow enough to support interpretation, and where it is not, the instrument exposes the interval so that the deviation can be declined rather than over-read.
Which domains fall on which side is not arbitrary.
The domain that separates the groups most cleanly, Writing Extent, is also the least reliable at the individual level, and both facts follow from one cause: it pools only a few, strongly correlated features, whose shared signal discriminates groups well but whose averaging cancels little of the reference sampling noise.
Group discriminability and individual reliability are therefore distinct virtues that can trade off.
The same two quantities, how many features a domain pools and how correlated they are, also set its intra-domain coherence and the dispersion of its calibrated scores, so reliability is not uniform across the profile but predictable from its construction.
The widest intervals, bounded here by the size of the healthy reference, are accordingly the ones a larger reference would most improve.

\subsection{Heterogeneity within the dysgraphia label}
The central rationale for a multi-axis measure, instead of a single label or a single global score, is that children who share the dysgraphia label do not share a single deficit.
In the 12-dimensional profile space, two dysgraphic children are on average as far apart as a dysgraphic child is from a typically-developing control (\S\ref{sec:res-e4het}).
This spread is multidimensional rather than a spread in how much children deviate overall.
With each child's overall deviation magnitude held fixed, profile shape (which domains deviate) varies almost as much as it does across all children, so two children with the same overall deviation can be flagged on entirely different processes (Appendix~\ref{app:shape}).
A single diagnostic label discards this structure, and a single global deviation score would collapse the differing shapes onto one number.
Only a per-domain profile preserves which processes a given child's difficulty implicates, which is the information an intervention would need to target.
We frame this heterogeneity as continuous and multidimensional, not as evidence for discrete dysgraphia subtypes, which a sample of this size cannot establish.

This within-label heterogeneity is the complement of the group-level phenotype.
The known-groups analysis identifies the processes on which dysgraphic children deviate on average, whereas the individual profiles show that the processes a given child actually deviates on are not fixed by that average, since a child may be flagged on a domain that carries no group-level effect while remaining typical on several that do.
A domain's inability to separate the groups therefore does not render it uninformative for an individual; it means only that group membership does not predict deviation there.
The profile is accordingly read per child rather than as a template of the group phenotype.
Part of this variation is structured rather than idiosyncratic.
Children carrying a dyslexia diagnosis alongside dysgraphia show amplified deviation on the temporal-organization and path-geometry domains relative to the dysgraphia-only subgroup.
The comorbid subgroup is small, but the observation illustrates the kind of profile structure, here aligned with a comorbidity, that a multidimensional measure can surface and a single label cannot.

Whether a given individual deviation reflects genuine difficulty or ordinary writing-style variation is a question of interpretation we take up next (\S\ref{sec:individual-limits}).

\subsection{Limits of individual-level interpretation}
\label{sec:individual-limits}
The known-groups evidence is a claim about groups, and it should not be read as a claim that any single domain separates individual children.
Even the strongest domain, D08~Writing~Extent, has a group effect of $\delta=+0.51$; because Cliff's $\delta$ is an ordinal-dominance probability, this corresponds to a healthy child scoring higher than a dysgraphic child on roughly one in four cross-group pairs ($P(\text{healthy}>\text{dysgraphic})\approx0.25$).
Substantial overlap of the two distributions is intrinsic to the effect sizes we report, and it is why the instrument is framed as a positional measure rather than a per-child classifier.

Defining the ``typical range'' as the empirical 5th--95th percentile band of the verified-typical reference on each domain, we find that $67\%$ of typical children already fall outside the band on at least one of the twelve domains (mean $1.31$ domains), against $67\%$ and mean $1.79$ domains for dysgraphic children.
The dichotomized risk ratio at $\geq$1 deviating domain is $\approx0.99$, and the groups separate only in the tail ($\geq$3 domains: $15\%$ vs.\ $29\%$; Fig.~\ref{fig:s-groupvsind}A).
The same reference-based measure that produces this heavy individual overlap also reproduces the group-level known-groups signal (Fig.~\ref{fig:s-groupvsind}B), so the effect is real at the group level and weak as an individual dichotomizer at the same time.
This is the expected behavior of a normative-reference instrument, in which about $5\%$ of typically developing children exceed the 95th height percentile of a growth standard without being pathological \citep{cole1990lms}.
A ``typical'' child who deviates on a domain is therefore best understood in one of three non-exclusive ways.
The deviation may be normal individual style variation, which is the positional reading above and is most likely on domains with no group-level effect, where a healthy child can deviate as much as a dysgraphic one.
It may be reference-sampling uncertainty, which the per-child bootstrap CI quantifies and which is widest on the high-coherence domains such as D08.
Or it may be genuine sub-clinical variation, that is, real handwriting difficulty below the diagnostic threshold, since a counselor-assigned label may classify a mildly affected child as typical.
Appropriate use follows from this: read per-domain position against a verified-typical reference rather than a category, read multi-domain patterns rather than any single axis, treat the bootstrap CI as a signal of when individual interpretation is unsafe, and generate hypotheses for follow-up rather than diagnoses.
An elevated per-child deviation rate is not a diagnostic rate, and clinical meaning requires external corroboration and professional judgment.

\subsection{Practical use: an open reference implementation}
Because nothing is fitted at request time and all calibrated state lives in a versioned, immutable artifact (\S\ref{sec:system}), a profile can be recomputed exactly, and scores obtained against the same artifact sit on a common scale.
That is what would allow separate studies to be compared rather than only described, and it is the property the released implementation is meant to supply.

\subsection{Limitations}
Individual-level reliability is bounded by the size of the healthy reference: intervals narrow approximately as $1/\sqrt{N}$, so a larger verified-typical sample is the direct remedy for the domains that are least precise here.

The reference is language- and script-specific.
All features are computed on sentence-level Czech writing (Latin script with diacritics).
Because a domain deviation is a joint motor--orthographic--linguistic quantity, the verified-typical reference is valid only for children writing Latin-based Czech sentences, and so is the open reference implementation, which ships only the Czech reference artifact.
Applying it to other languages or scripts (for example English, Arabic, Chinese, or Korean) is unestablished, and a replication within a closely related language would not by itself test cross-script generality.
The instrument should not be applied to non-Czech writing without re-establishing the reference on an appropriate verified-typical sample.
Even within Czech, the reference is built from a single cohort and is therefore cohort-specific until replicated.

External-criterion validity is a boundary of the present work.
Beyond the known-groups contrast, the DiaGraMo battery contains no objectively scored, reliability-verified handwriting-quality criterion, so we do not report convergent validity against an external handwriting measure.
Its only handwriting criterion, the HPSQ-C, is a parent/teacher report whose internal consistency could not be estimated in this cohort.
The cognitive scales the battery does contain, namely phonological awareness, visuospatial recall, and verbal and quantitative reasoning, are not handwriting-independent, because handwriting is a multiply determined act that draws on visuomotor and linguistic processes and frequently co-occurs with dyslexia.
Treating them as discriminant criteria would therefore be theoretically mis-specified.

The group label is a counselor judgment rather than a score on a standardized diagnostic instrument, so the known-groups contrast is anchored to a criterion whose own measurement properties are unknown, and some of the modest effects may reflect label noise as much as a small true difference.
The cohort also contains no validated motor-severity scale, so the profile gallery stratifies children by an external parent-report legibility axis purely to select illustrative cases non-circularly, and demonstrates profile heterogeneity rather than a severity gradient.
The vocabulary's content validity, finally, rests on structural coherence and literature grounding, not on an independent expert-panel rating, which we did not conduct.

Several further boundaries follow from the study's scope.
Each child was recorded once, so the bootstrap quantifies how much a score depends on which children constitute the reference, but not how much it would move if the same child wrote the sentence again on another day.
The protocol--age confound is mitigated but not eliminated, and it is the three sign-unstable domains of \S\ref{sec:res-s2} that remain exposed to it, which is why we draw no directional conclusion from them.
True on-surface pauses, meaning pen-down hesitations, are not measured: the frozen extractor operationalizes ``pause'' as the duration of an in-air segment, so those features belong to D05~In-Air~Behavior.
The recordings do carry what a pen-down pause measure would need, and adding one is a matter of choosing velocity and duration thresholds, but that would alter the frozen feature set and is left to a revised vocabulary.

\subsection{Future work}
The most important direction is an in-the-loop study of clinical utility, placing the profile in a real clinical or educational workflow and testing whether it improves practitioners' decisions or children's outcomes, which would turn the illustrative usage scenarios into evaluated ones.
The rest answer the limitations above in turn.

A larger verified-typical reference would narrow every interval, and would help most on the domains that are least precise here.
Reference artifacts fitted on other languages and scripts would extend the instrument beyond Czech, and replication on an independent cohort would show whether the group-level pattern transports beyond this sample.
Convergent validity should be tested against a reliable, objectively scored handwriting-quality instrument rather than a parent-report screen, and the group contrast against a standardized diagnostic instrument rather than a counselor-assigned label.
Discriminant validity should be reframed to respect the multiply determined nature of handwriting: rather than seeking cognitive constructs assumed to be orthogonal to it, the appropriate test is one of incremental validity, whether the domain deviations predict a handwriting criterion beyond general cognitive ability, controlling for the visuospatial, phonological, and other cognitive skills with which handwriting shares variance.
A calibrated motor-severity anchor could replace the illustrative-only legibility axis, and an independent expert panel could rate the vocabulary's content validity.
A test--retest cohort would supply the temporal stability the present design cannot, and a pre/post cohort would test responsiveness: whether a follow-up profile scored against the same frozen reference detects change on the domains an intervention targets, where a binary screen would return the same label at both time points.
A cohort in which protocol is not confounded with age would settle the domains that remain exposed to that confound.
A refined extractor could add a true on-surface pause domain, and a complementary graphomotor-primitive vocabulary could extend coverage to the non-linguistic tasks left out here.

Whether the continuous heterogeneity we document partitions into discrete subtypes is a further, stronger question for larger pre-registered samples.

%% file: sections/5_conclusion.tex
\section{Conclusion}
We have introduced and validated an open instrument for profiling handwriting-process deviations in children with developmental dysgraphia.
It pairs a literature-only, twelve-domain vocabulary over 136 online-handwriting features with an age- and sex-adjusted normative reference, turning a child's pen trajectory into a twelve-axis deviation profile against verified-typical peers, with explicit per-child uncertainty.

The contribution is the validated method rather than any single artifact.
Four measurement properties support the profile as a usable measure: structural coherence, a calibrated and parsimonious reference, known-groups construct validity, and individual-level reliability.
Because the feature-to-domain mapping was frozen before any cohort analysis, none of these rests on a model shaped by the results it produced.
What the instrument returns is a position against a healthy reference rather than a diagnostic label.

To make the validated method reusable, we openly release every artifact: the vocabulary, the analysis code, and an open reference implementation that computes the same profile for new participants.
In moving handwriting assessment from a single label or global score to a graded, multidimensional profile, the instrument gives researchers a way to decompose the handwriting process into interpretable dimensions and to situate an individual child against a normative reference.
Whether acting on such profiles improves outcomes is a question of clinical utility that we leave to future in-the-loop study.

%% file: sections/back_matter.tex

\section*{Declarations}

\bmhead{Funding}
This work was supported by Hi! PARIS and the ANR/France 2030 program [grant number ANR-23-IACL-0005].

\bmhead{Competing interests}
The authors have no competing interests to declare that are relevant to the
content of this article.

\bmhead{Ethics approval}
This study is a secondary analysis of the publicly available, de-identified DiaGraMo dataset (CC-BY-4.0; \citealp{zvonvcakova2026multimodal}). 
Because the present work analyses only fully de-identified, openly licensed data and involves no new data collection or participant contact, no additional ethics approval was required.

\bmhead{Consent for publication}
Not applicable. The article contains no individually identifiable information; all records are pseudonymous and no participant can be identified from any figure or table.

\bmhead{Availability of data and materials}
The DiaGraMo dataset analyzed in this study is openly available on Zenodo (\url{https://doi.org/10.5281/zenodo.18299327}; CC-BY-4.0). 
The age- and sex-adjusted verified-typical reference artifact derived here (a healthy-only feature matrix with fitted reference coefficients) is released with the reference implementation at \url{https://github.com/jihyunmun/handwriting-process-profiling}; because it is derived from DiaGraMo, it is redistributed under CC-BY-4.0 with attribution to that dataset.
The verbatim items and scoring manuals of the standardized tests (WJ-IV, BACH, RCFT) are copyright-protected and are \emph{not} redistributed; only derived numeric scores are used.

\bmhead{Code availability}
The vocabulary (\texttt{vocabulary\_v1.csv}), the full analysis code, and the open reference implementation are available at \url{https://github.com/jihyunmun/handwriting-process-profiling} under a permissive (MIT) license and archived on Zenodo (\url{https://doi.org/10.5281/zenodo.22746861}); each reported result is reproducible from a documented script entry-point.

\bmhead{Authors' contributions}
J.M.: conceptualization, methodology, software, validation, formal analysis,
writing -- original draft, writing -- review \& editing.
M.A.E.-Y.: conceptualization, supervision, writing -- review \& editing.
Both authors read and approved the final manuscript.

\bmhead{Use of large language models}
During the preparation of this work, the authors used large language model assistants for language editing and readability improvement.
No text was generated autonomously for the scientific content; the authors reviewed and edited all material and take full responsibility for the content of the article.

\bmhead{Open Practices Statement}
The data (DiaGraMo) are openly available on Zenodo (\url{https://doi.org/10.5281/zenodo.18299327}).
The materials and analysis code (the vocabulary, the analysis scripts, and the reference implementation) are available at \url{https://github.com/jihyunmun/handwriting-process-profiling} and are archived on Zenodo (\url{https://doi.org/10.5281/zenodo.22746861}).
None of the analyses reported here was preregistered; the 12-domain handwriting-process vocabulary was derived from the literature alone and fixed before any analysis of the cohort.

%% file: sections/appendix.tex

\clearpage
\begin{appendices}

\setcounter{table}{0}
\renewcommand{\thetable}{S\arabic{table}}
\setcounter{figure}{0}
\renewcommand{\thefigure}{S\arabic{figure}}


\section{Reference-framework supporting detail}\label{app:refdetail}
This appendix collects the per-domain detail behind three reference-framework checks summarized in the main text: model-variant robustness, self-inclusion leakage, and effect-size stability.

\paragraph{Model-variant robustness.}
We recomputed the per-domain known-groups Cliff's $\delta$ under four reference variants: the default leave-one-subject-out (LOSO) refit, an in-sample fit, a fit adding a protocol indicator as a covariate, and a sex-stratified fit (Table~\ref{tab:s-robust}).
The largest change relative to the LOSO default is $0.030$ (D05, protocol-covariate variant), and all effect directions are preserved.

\begin{table}[h]
\centering
\caption{Model-variant robustness: per-domain Cliff's $\delta$ under four
reference variants. The last column is the maximum absolute deviation of the
three alternative variants from the LOSO default (maximum across domains
$=0.030$).}
\label{tab:s-robust}
\small
\begin{tabular}{@{}lccccc@{}}
\toprule
Domain & LOSO & In-sample & Protocol-cov. & Sex-strat. & Max $|\Delta|$ \\
\midrule
D01 Writing Speed       & $-0.006$ & $-0.010$ & $-0.017$ & $-0.018$ & 0.012 \\
D02 Speed Consistency   & $+0.092$ & $+0.092$ & $+0.090$ & $+0.093$ & 0.002 \\
D03 Movement Smoothness & $+0.135$ & $+0.138$ & $+0.134$ & $+0.143$ & 0.009 \\
D04 Temporal Org.       & $+0.477$ & $+0.483$ & $+0.483$ & $+0.490$ & 0.012 \\
D05 In-Air Behavior     & $+0.358$ & $+0.364$ & $+0.389$ & $+0.374$ & 0.030 \\
D06 Stroke Timing       & $-0.059$ & $-0.058$ & $-0.045$ & $-0.055$ & 0.014 \\
D07 Stroke Size         & $+0.154$ & $+0.156$ & $+0.153$ & $+0.152$ & 0.002 \\
D08 Writing Extent      & $+0.510$ & $+0.517$ & $+0.521$ & $+0.516$ & 0.011 \\
D09 Path Geometry       & $+0.094$ & $+0.094$ & $+0.088$ & $+0.089$ & 0.006 \\
D10 Pressure Control    & $+0.148$ & $+0.155$ & $+0.147$ & $+0.156$ & 0.008 \\
D11 Pen Tilt            & $+0.187$ & $+0.190$ & $+0.189$ & $+0.198$ & 0.010 \\
D12 Pen Rotation        & $+0.402$ & $+0.402$ & $+0.403$ & $+0.409$ & 0.007 \\
\botrule
\end{tabular}
\end{table}

\paragraph{Self-inclusion leakage.}
For each healthy child we compare the domain deviation magnitude obtained when the child is included in the reference that scores them (in-sample) against the leave-one-subject-out (LOSO) value used throughout (Table~\ref{tab:s-leak}).
Self-inclusion deflates the mean healthy $|\overline{z}|$ by a median of only $0.022$ (at most $0.026$, D08), so leakage is negligible and LOSO removes it by construction.

\begin{table}[t]
\centering
\caption{Self-inclusion leakage: per-domain mean healthy $|\overline{z}|$ under
the LOSO refit vs.\ an in-sample reference, and their difference. A positive
difference is the deflation removed by LOSO (median $0.022$, max $0.026$).}
\label{tab:s-leak}
\small
\begin{tabular}{@{}lccc@{}}
\toprule
Domain & LOSO $\langle|\overline{z}|\rangle$ & In-sample $\langle|\overline{z}|\rangle$ & Difference \\
\midrule
D01 Writing Speed       & 0.337 & 0.318 & $+0.019$ \\
D02 Speed Consistency   & 0.423 & 0.403 & $+0.021$ \\
D03 Movement Smoothness & 0.251 & 0.228 & $+0.023$ \\
D04 Temporal Org.       & 0.359 & 0.346 & $+0.013$ \\
D05 In-Air Behavior     & 0.324 & 0.303 & $+0.022$ \\
D06 Stroke Timing       & 0.421 & 0.402 & $+0.019$ \\
D07 Stroke Size         & 0.433 & 0.410 & $+0.024$ \\
D08 Writing Extent      & 0.649 & 0.623 & $+0.026$ \\
D09 Path Geometry       & 0.483 & 0.461 & $+0.023$ \\
D10 Pressure Control    & 0.264 & 0.254 & $+0.010$ \\
D11 Pen Tilt            & 0.401 & 0.379 & $+0.022$ \\
D12 Pen Rotation        & 0.466 & 0.444 & $+0.022$ \\
\botrule
\end{tabular}
\end{table}

\paragraph{Effect-size stability.}
We bootstrap-resampled the healthy reference ($B=1000$); in each resample we recomputed the twelve per-domain group effect sizes and re-ranked the domains by $|\delta|$.
The Spearman correlation between each resample's ranking and the point-estimate ranking has median $0.979$ (IQR $0.028$), i.e.\ the ordering of the domains by effect size is essentially invariant to which healthy children constitute the reference.
Table~\ref{tab:s-stab} reports the per-domain point estimate with its bootstrap 95\% CI and CI width.

\begin{table}[t]
\centering
\caption{Effect-size stability: per-domain group Cliff's $\delta$ point estimate
with bootstrap 95\% CI and CI width ($B=1000$). Median rank correlation of the
domain ordering across resamples $=0.979$ (IQR $0.028$).}
\label{tab:s-stab}
\small
\begin{tabular}{@{}lccc@{}}
\toprule
Domain & Cliff's $\delta$ & 95\% CI & CI width \\
\midrule
D01 Writing Speed       & $-0.006$ & [$-0.070$, $+0.014$] & 0.084 \\
D02 Speed Consistency   & $+0.092$ & [$+0.064$, $+0.118$] & 0.053 \\
D03 Movement Smoothness & $+0.135$ & [$+0.050$, $+0.224$] & 0.174 \\
D04 Temporal Org.       & $+0.477$ & [$+0.457$, $+0.498$] & 0.040 \\
D05 In-Air Behavior     & $+0.358$ & [$+0.324$, $+0.397$] & 0.073 \\
D06 Stroke Timing       & $-0.059$ & [$-0.086$, $-0.032$] & 0.055 \\
D07 Stroke Size         & $+0.154$ & [$+0.127$, $+0.175$] & 0.048 \\
D08 Writing Extent      & $+0.510$ & [$+0.486$, $+0.526$] & 0.039 \\
D09 Path Geometry       & $+0.094$ & [$+0.084$, $+0.105$] & 0.021 \\
D10 Pressure Control    & $+0.148$ & [$+0.134$, $+0.171$] & 0.037 \\
D11 Pen Tilt            & $+0.187$ & [$+0.158$, $+0.216$] & 0.057 \\
D12 Pen Rotation        & $+0.402$ & [$+0.347$, $+0.417$] & 0.070 \\
\botrule
\end{tabular}
\end{table}

\clearpage
\section{Feature-level effect sizes}\label{app:feature}
As a check that domain aggregation preserves rather than obscures signal, we repeated the known-groups contrast at the level of individual features.
Of the 136 features, 58 are individually Benjamini--Hochberg-significant at $q<0.05$ (corrected over all 136 features); the strongest is \texttt{jerk\_cv} (D03 Movement Smoothness, $\delta=-0.612$).
Table~\ref{tab:s-feature} lists the 20 features with the largest $|\delta|$.
The significant features fall in the domains that carry the group signal (D04, D05, D08, and D12) together with the two near-significant domains (D03, D07), so the domain-level and feature-level analyses implicate the same processes.
The single strongest feature nonetheless sits in D03~Movement~Smoothness, which only reaches near-significance once its features are averaged.

\begin{table}[h]
\centering
\caption{Feature-level known-groups effect sizes (dysgraphic vs.\ healthy):
the 20 features with the largest $|\delta|$, out of 136. Cliff's $\delta$ with
Benjamini--Hochberg $q$ (corrected over all 136 features). 58/136 features are
significant at $q<0.05$.}
\label{tab:s-feature}
\small
\begin{tabular}{@{}llcc@{}}
\toprule
Feature & Domain & Cliff's $\delta$ & BH $q$ \\
\midrule
\texttt{jerk\_cv}                     & D03 & $-0.612$ & $6.4\mathrm{e}{-}15$ \\
\texttt{writing\_width}               & D08 & $+0.532$ & $2.1\mathrm{e}{-}11$ \\
\texttt{in\_air\_stroke\_duration\_total} & D05 & $+0.519$ & $5.2\mathrm{e}{-}11$ \\
\texttt{in\_air\_time\_ms}            & D04 & $+0.486$ & $8.7\mathrm{e}{-}10$ \\
\texttt{writing\_height}              & D08 & $+0.466$ & $4.4\mathrm{e}{-}09$ \\
\texttt{vertical\_distance}           & D08 & $+0.461$ & $5.8\mathrm{e}{-}09$ \\
\texttt{stroke\_height\_median}       & D07 & $+0.456$ & $8.2\mathrm{e}{-}09$ \\
\texttt{jerk\_mean}                   & D03 & $+0.445$ & $1.8\mathrm{e}{-}08$ \\
\texttt{total\_distance}              & D08 & $+0.439$ & $2.7\mathrm{e}{-}08$ \\
\texttt{stroke\_height\_max}          & D07 & $+0.426$ & $6.9\mathrm{e}{-}08$ \\
\texttt{stroke\_length\_median}       & D07 & $+0.408$ & $2.7\mathrm{e}{-}07$ \\
\texttt{acceleration\_cv}             & D03 & $-0.402$ & $4.0\mathrm{e}{-}07$ \\
\texttt{nca}                          & D03 & $+0.396$ & $6.1\mathrm{e}{-}07$ \\
\texttt{stroke\_width\_median}        & D07 & $+0.394$ & $6.3\mathrm{e}{-}07$ \\
\texttt{azimuth\_iqr}                 & D12 & $+0.391$ & $7.5\mathrm{e}{-}07$ \\
\texttt{azimuth\_std}                 & D12 & $+0.389$ & $8.1\mathrm{e}{-}07$ \\
\texttt{ncv}                          & D03 & $+0.383$ & $1.2\mathrm{e}{-}06$ \\
\texttt{on\_surface\_time\_ms}        & D04 & $+0.378$ & $1.7\mathrm{e}{-}06$ \\
\texttt{horizontal\_distance}         & D08 & $+0.369$ & $3.0\mathrm{e}{-}06$ \\
\texttt{azimuth\_cv}                  & D12 & $+0.359$ & $5.7\mathrm{e}{-}06$ \\
\botrule
\end{tabular}
\end{table}

\clearpage
\section{Comorbidity stratification}\label{app:comorbid}
Because 29 of the 147 dysgraphic children also carry a dyslexia diagnosis, we stratified the known-groups contrast into dysgraphia-only ($N=115$) and dysgraphia-plus-dyslexia ($N=29$) subgroups, each against the healthy reference (Table~\ref{tab:s-comorbid}).
The dysgraphia-only subgroup reproduces the pooled pattern; the comorbid subgroup shows amplified deviation on D04 Temporal Organization ($\delta=+0.621$ vs.\ $+0.440$) and D09 Path Geometry ($+0.268$ vs.\ $+0.039$).
Given the small comorbid $N$, this is reported as an exploratory, descriptive observation without significance testing.

\begin{table}[h]
\centering
\caption{Comorbidity stratification: per-domain Cliff's $\delta$ against the
healthy reference for the dysgraphia-only subgroup ($N=115$) and the
dysgraphia-plus-dyslexia subgroup ($N=29$), and their difference. Exploratory
and descriptive; no significance testing.}
\label{tab:s-comorbid}
\small
\begin{tabular}{@{}lccc@{}}
\toprule
Domain & Dys-only & Dys+dyslexia & Difference \\
\midrule
D01 Writing Speed       & $-0.014$ & $+0.014$ & $+0.024$ \\
D02 Speed Consistency   & $+0.109$ & $+0.043$ & $-0.054$ \\
D03 Movement Smoothness & $+0.135$ & $+0.130$ & $-0.003$ \\
D04 Temporal Org.       & $+0.440$ & $+0.621$ & $+0.219$ \\
D05 In-Air Behavior     & $+0.353$ & $+0.381$ & $+0.018$ \\
D06 Stroke Timing       & $-0.056$ & $-0.045$ & $+0.036$ \\
D07 Stroke Size         & $+0.144$ & $+0.181$ & $+0.048$ \\
D08 Writing Extent      & $+0.510$ & $+0.522$ & $+0.038$ \\
D09 Path Geometry       & $+0.039$ & $+0.268$ & $+0.238$ \\
D10 Pressure Control    & $+0.149$ & $+0.146$ & $-0.017$ \\
D11 Pen Tilt            & $+0.188$ & $+0.140$ & $-0.043$ \\
D12 Pen Rotation        & $+0.428$ & $+0.330$ & $-0.077$ \\
\botrule
\end{tabular}
\end{table}

\clearpage
\section{Multidimensionality of within-label heterogeneity}\label{app:shape}
\begin{figure*}[t]
\centering
\includegraphics[width=\textwidth]{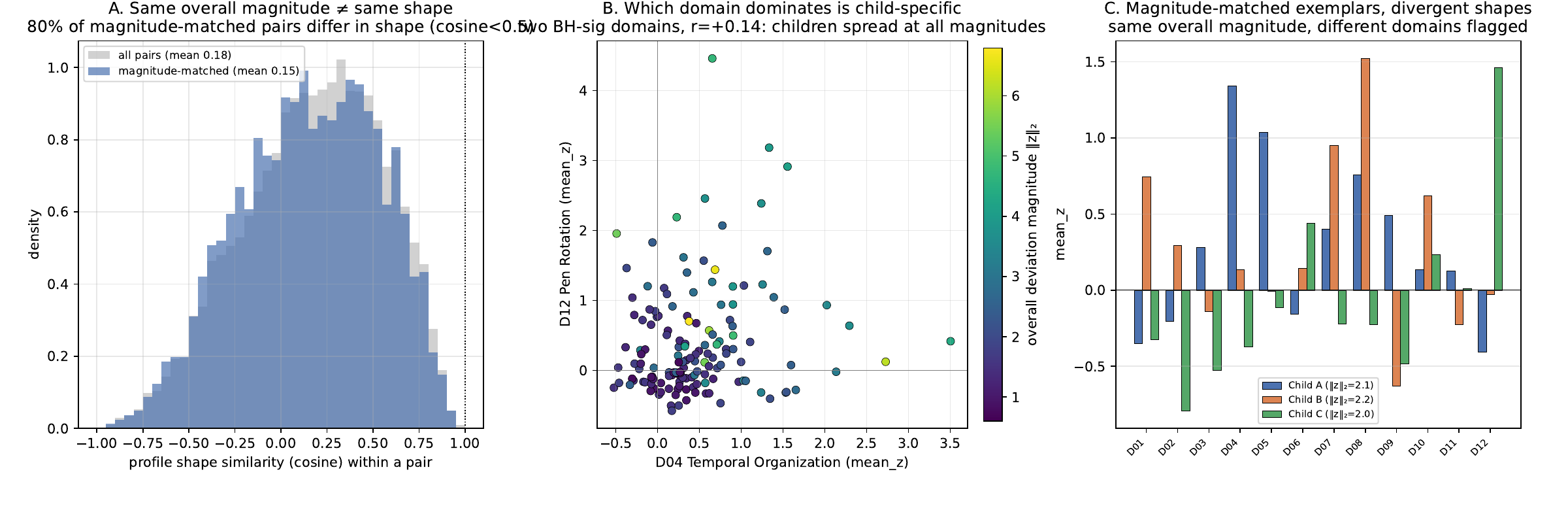}
\caption{Within-label heterogeneity is multidimensional (profile shape), not
merely overall deviation magnitude (dysgraphic children, $N=145$).
\textbf{(A)} Pairwise profile-shape similarity (cosine, magnitude removed) for
magnitude-matched pairs vs.\ all pairs: the distributions nearly coincide
(mean $0.15$ vs.\ $0.18$) and $80\%$ of magnitude-matched pairs differ clearly in
shape (cosine ${<}0.5$).
\textbf{(B)} Two known-groups-significant but weakly correlated domains
(D04 Temporal Organization vs.\ D12 Pen Rotation, $r=+0.14$), colored by overall
deviation magnitude $\lVert z\rVert_2$: at every magnitude level children spread
across the plane, so which domain dominates is child-specific.
\textbf{(C)} Three children from one narrow magnitude band
($\lVert z\rVert_2\approx2$) with divergent profiles: the same overall magnitude
flags different domains. Neither a binary label nor a single aggregate deviation
score can substitute for the 12-domain profile; the variation is continuous, and
no discrete-subtype claim is made. See Appendix~\ref{app:shape} for definitions
and a full reading.}
\label{fig:s-shape}
\end{figure*}

This appendix expands the Aim~4 finding (\S\ref{sec:res-e4het}) that the within-dysgraphia heterogeneity is multidimensional rather than a one-dimensional spread in how much children deviate, and details Fig.~\ref{fig:s-shape}. It uses dysgraphic children only ($N=145$; two children with an implausible D02~Speed~Consistency value, the least-reliable domain, are removed so the decomposition is not driven by a single-domain artifact).

\paragraph{Definitions.}
Each child is summarized by a 12-dimensional profile of domain deviation scores. We decompose a profile into two independent descriptors. Its \emph{overall deviation magnitude} is the Euclidean length $\lVert z\rVert_2$ of the profile, a single descriptive number for how far the whole profile sits from the typical-child origin. It is \emph{not} a validated clinical severity rating: this instrument makes no severity-calibration claim, and we use ``magnitude'' throughout rather than ``severity'' to avoid that reading. Its \emph{shape} is the unit-length direction $z/\lVert z\rVert_2$: which domains deviate, and in what relative proportions, with magnitude divided out. The \emph{shape similarity} of two children is the cosine of the angle between their profiles ($1$ = identical shape, $0$ = unrelated, ${<}0$ = opposite pattern); being magnitude-invariant, it isolates ``which domains'' from ``how much''.

\paragraph{Reading Fig.~\ref{fig:s-shape}.}
\emph{Panel A} contrasts the distribution of pairwise shape similarity for magnitude-matched pairs (children whose overall magnitudes lie within $0.25$ $z$-units) against that for all pairs. If overall magnitude determined shape, matched pairs would pile up near cosine $1$; instead the two distributions almost coincide (mean $0.15$ vs.\ $0.18$) and $80\%$ of matched pairs differ clearly in shape (cosine ${<}0.5$), so knowing that two children share an overall magnitude is nearly uninformative about their profile shape.
\emph{Panel B} plots two known-groups-significant but weakly correlated domains (D04 Temporal Organization vs.\ D12 Pen Rotation; within-group $r=+0.14$), one point per child, colored by overall magnitude. At every magnitude level children spread across the plane, some D04-dominant and some D12-dominant, so which domain drives a child's deviation is child-specific rather than fixed by overall magnitude or by the group. Across the five significant domains the mean absolute within-group intercorrelation is $0.22$ (a pure magnitude axis would give $\approx 1$), and $64\%$ of children have their largest deviation on a domain other than the group-modal D08 Writing Extent.
\emph{Panel C} shows three children drawn from a single narrow magnitude band ($\lVert z\rVert_2\approx 2$) whose profiles nonetheless diverge: one D04/D05-dominant, one D08-dominant, one D12-dominant, the same ``how much'' pointing to different domains.

\paragraph{Scope.}
These analyses characterize the heterogeneity as continuous and multidimensional; they are not a subtype analysis. Whether the dysgraphic children further partition into discrete subtypes is a separate, stronger question that a sample of this size cannot settle, so we make no dysgraphia-subtype claim and leave subtype discovery to larger, pre-registered samples.

\clearpage
\section{Profile-gallery selection}\label{app:gallery}
The profiles discussed in the Aim~4 profile gallery (\S\ref{sec:res-e4b}) were chosen by the deterministic rule of Section~\ref{sec:validation-design}; Table~\ref{tab:e4b} records the resulting subjects (by display code: DG1--DG4 dysgraphic, HC1--HC4 matched healthy; participant identifiers are withheld, and the released code reproduces the selection), their HPSQ-C legibility subscale scores, and their matched healthy controls for reproducibility.

\begin{table}[h]
\centering
\caption{Aim~4 profile-gallery selection. One dysgraphic
child per HPSQ-C legibility quartile (leg $=$ subscale score), each matched to
a same-sex healthy child within 0.5~yr. Quartiles index parent-reported
concern, not motor severity.}
\label{tab:e4b}
\small
\begin{tabular}{@{}lllc@{}}
\toprule
Quartile & Dysgraphic (leg) & Matched healthy & Age \\
\midrule
Q1 (lowest concern)  & DG1 (1) & HC1 & 10.2 \\
Q2                   & DG2 (3)  & HC2 & 10.2 \\
Q3                   & DG3 (5) & HC3 & 9.9  \\
Q4 (highest concern) & DG4 (7) & HC4 & 9.3  \\
\botrule
\end{tabular}
\end{table}

\clearpage
\section{Hypothesis-generating interpretation guide}\label{app:interp}
Table~\ref{tab:s-interp} links each of the 12 domains to a possible reading of a deviation and a literature-suggested direction for targeted assessment.
Every entry uses strictly tentative (``may indicate'') wording; the cross-sectional design does not separate motor, orthographic, and linguistic mechanisms, deviations are described as handwriting-process deviations rather than motor deficits, and no treatment-efficacy claim is made.

\begin{table*}[h]
\centering
\caption{Per-domain hypothesis-generating interpretation guide. Readings and
directions are hypotheses for prospective validation, not diagnostic or
treatment-efficacy claims.}
\label{tab:s-interp}
\footnotesize
\begin{tabularx}{\textwidth}{@{}l X X@{}}
\toprule
Domain & Possible reading of a deviation (hypothesis-generating) & Literature-suggested direction (to be validated) \\
\midrule
D01 Writing Speed & May indicate unusually rapid writing (fluency ease or rushing) or slowed writing (possible motor-retrieval or planning difficulty). & May inform paced-writing protocols; targeted assessment needed to localize the mechanism. \\
D02 Speed Consistency & May indicate irregular writing rhythm, linked in the literature to attentional or motor-program instability. & May inform rhythm-based training; relation to symptoms should be verified. \\
D03 Movement Smoothness & May indicate less smooth or more segmented movement (motor-control difficulty or added cognitive load). & May inform smoothness-focused training (e.g., continuous tracing); mechanism requires follow-up. \\
D04 Temporal Organization & May indicate altered on-surface / in-air time allocation, reflecting planning, hesitation, or segmentation. & May inform fluency-oriented protocols; translation requires pre-specified follow-up. \\
D05 In-Air Behavior & May indicate altered pen-off-surface motion, linked to inter-stroke planning load, cognitive demand, or hesitation. & May inform inter-stroke planning / fluency tasks; true on-surface pauses are future work. \\
D06 Stroke Timing & May indicate more variable per-stroke duration, associated with motor-program instability. & May inform paced-stroke practice; relation to symptoms should be verified. \\
D07 Stroke Size & May indicate atypical stroke-size patterns, reflecting motor calibration or size regulation. & May inform size-targeted practice; causal attribution requires follow-up. \\
D08 Writing Extent & May indicate a larger or smaller overall writing product than age expectations (spatial-regulation difficulty). & May inform paper-boundary / spatial-awareness exercises; requires validation. \\
D09 Path Geometry & May indicate atypical trajectory curvature (motor-planning or visuomotor-integration differences). & Interpret cautiously: D09 coherence CI crossed zero; may span heterogeneous sub-constructs. \\
D10 Pressure Control & May indicate altered pressure regulation, associated with effort and discomfort. & May inform grip / pressure-feedback strategies; requires validated ergonomic assessment. \\
D11 Pen Tilt & May indicate atypical pen-holding altitude (grip or posture differences). & May inform ergonomic grip assessment; mechanism not identifiable from cross-sectional data. \\
D12 Pen Rotation & May indicate atypical pen-holding rotation (grip or hand-posture differences). & May inform ergonomic assessment; follow-up needed to separate grip from dynamic rotation. \\
\botrule
\end{tabularx}
\end{table*}

\clearpage
\section{Additional figures}\label{app:figs}

\begin{figure*}[h]
\centering
\includegraphics[width=\textwidth]{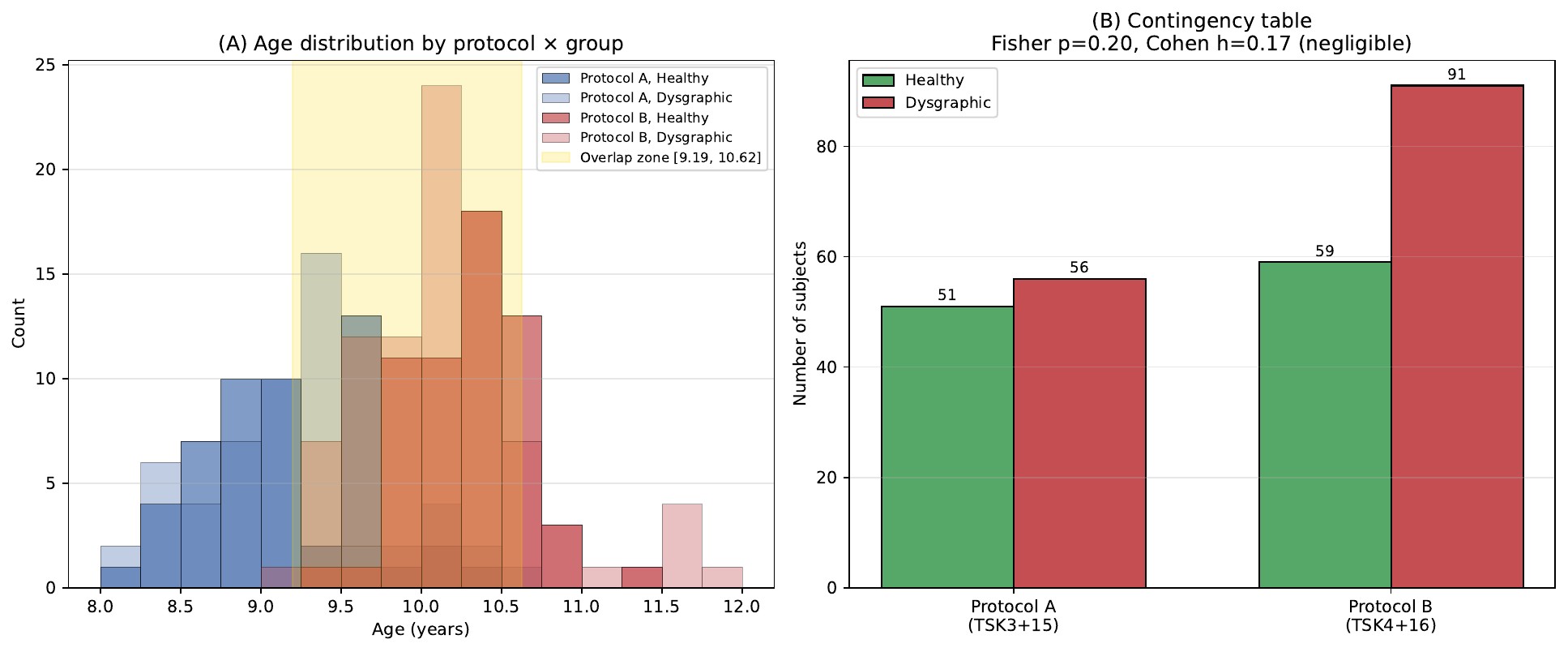}
\caption{Protocol--age confound disclosure. Protocol is age-stratified by the
DiaGraMo study design; within each protocol, healthy and dysgraphic children
are age-matched.}
\label{fig:s-protocol}
\end{figure*}

\begin{figure*}[h]
\centering
\includegraphics[width=\textwidth]{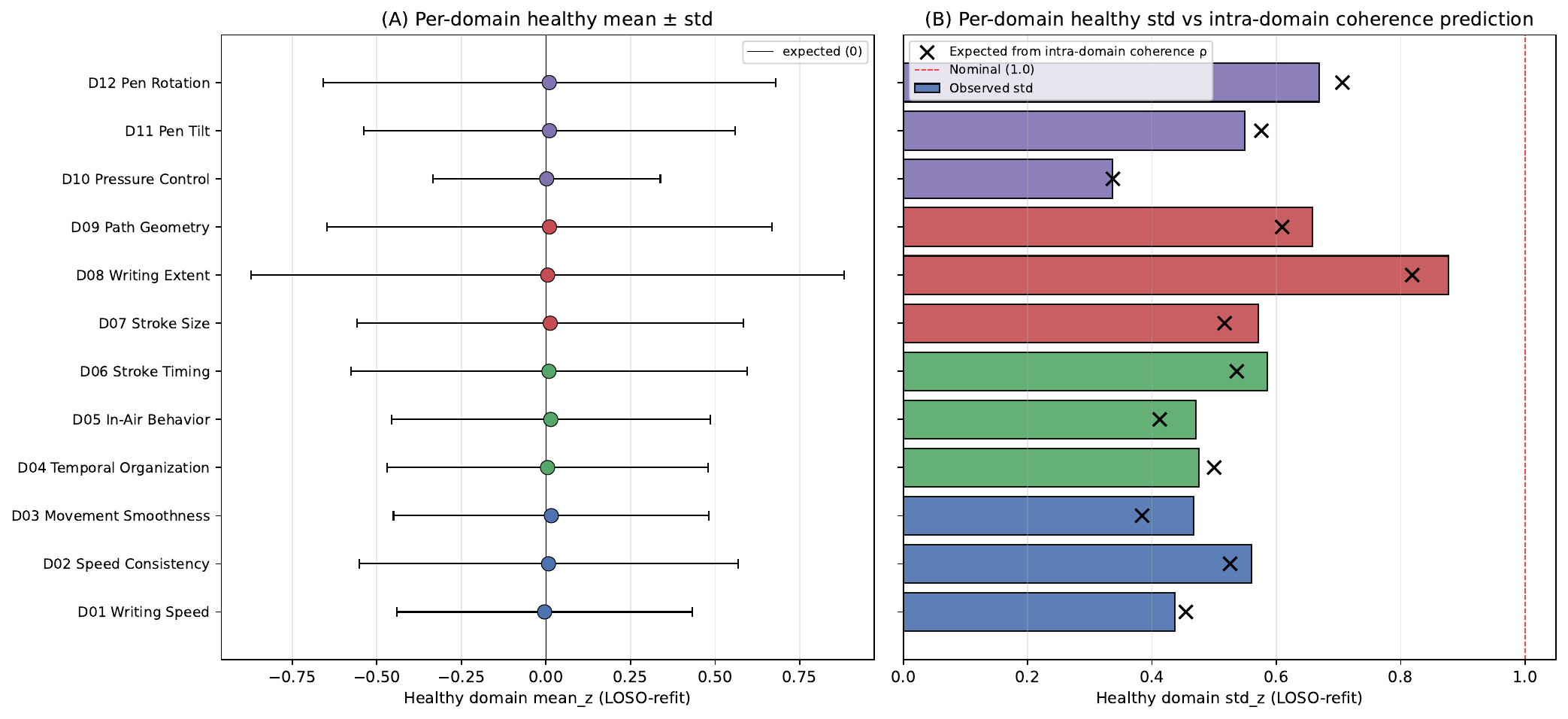}
\caption{Healthy-on-healthy calibration. Per-domain healthy means are near zero
(no systematic bias), and the observed healthy standard deviation tracks the
prediction $\sqrt{(1+(k-1)\rho)/k}$; sub-unit dispersion is the expected
consequence of aggregating correlated $z$-scores.}
\label{fig:s-calib}
\end{figure*}

\begin{figure*}[h]
\centering
\includegraphics[width=\textwidth]{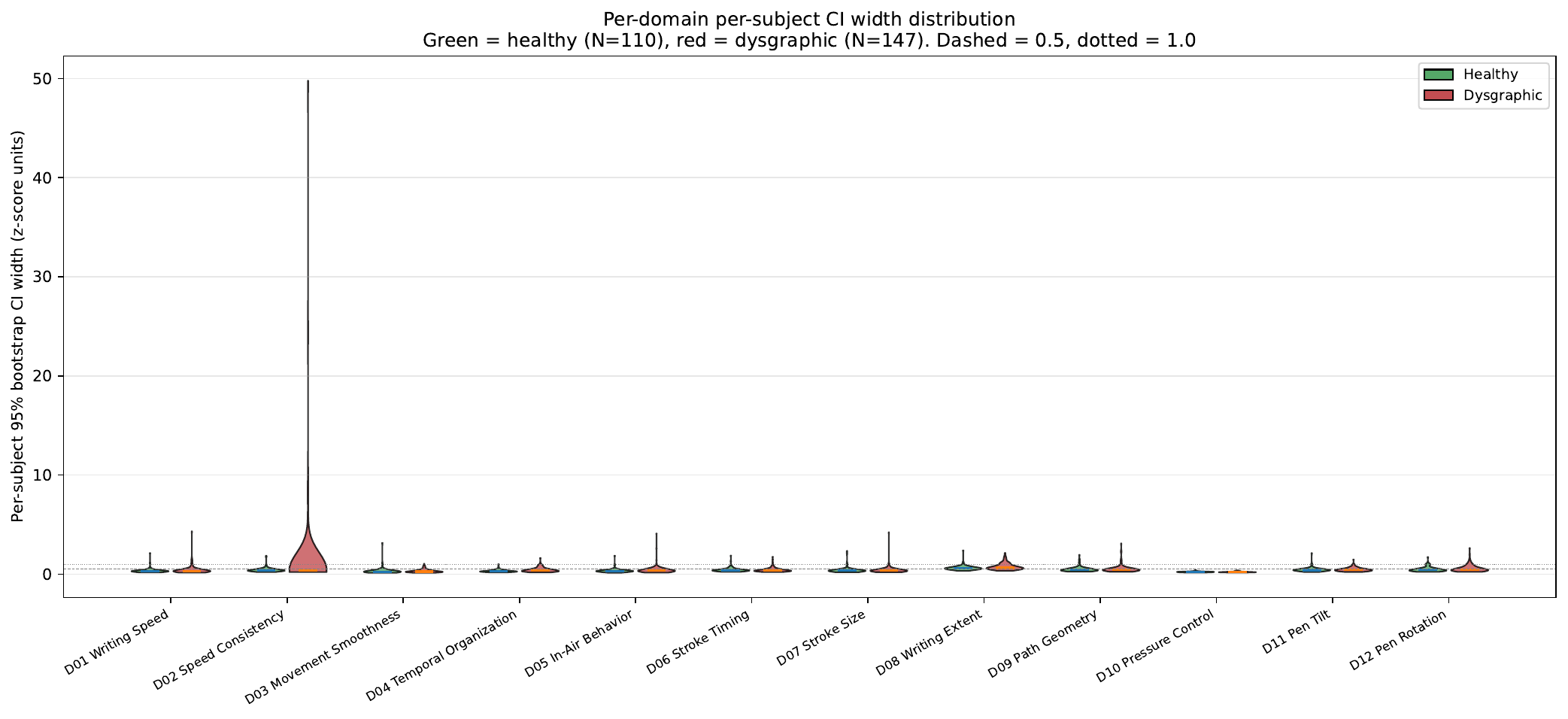}
\caption{Per-domain distribution of per-child 95\% bootstrap CI widths (healthy
vs.\ dysgraphic). Reference lines at 0.5 and 1.0 z-units; 96.5\% of
child~$\times$~domain pairs fall below 1.0.}
\label{fig:s-ci}
\end{figure*}

\begin{figure*}[h]
\centering
\includegraphics[width=\textwidth]{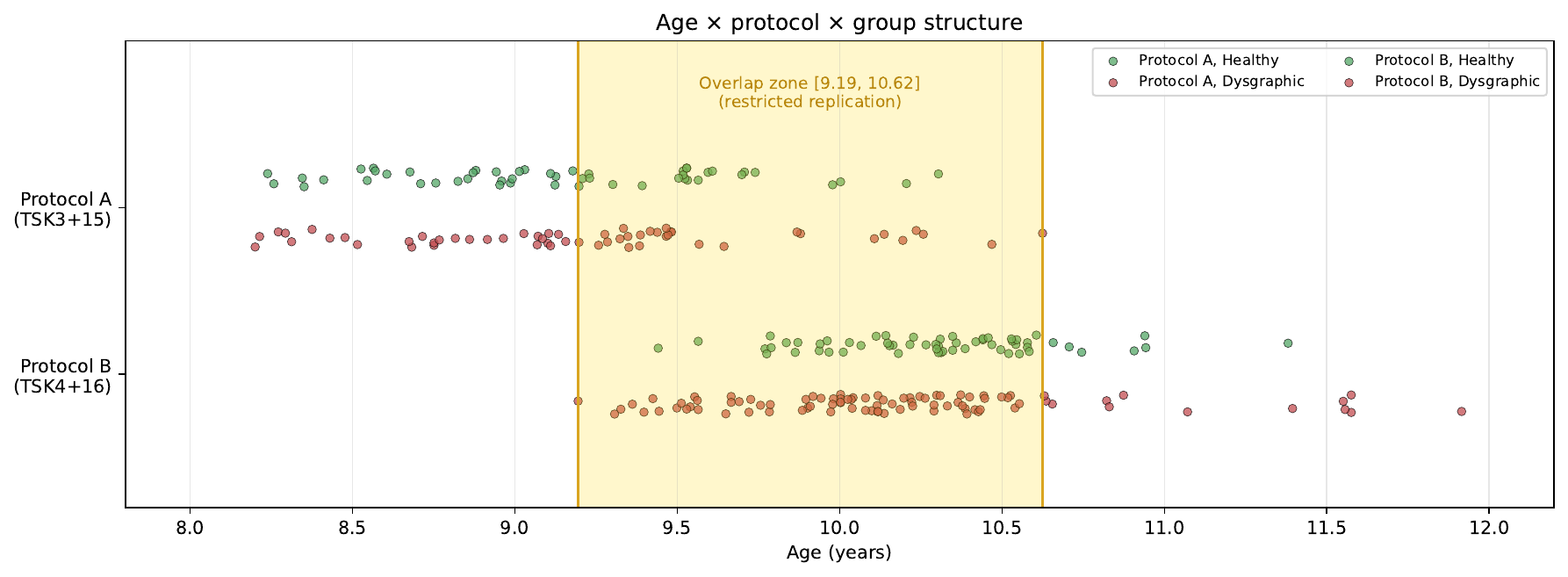}
\caption{Age~$\times$~protocol~$\times$~group structure with the age-overlap
band in which both protocols coexist, used for the restricted-replication
robustness check.}
\label{fig:s-overlap}
\end{figure*}

\begin{figure*}[h]
\centering
\includegraphics[width=\textwidth]{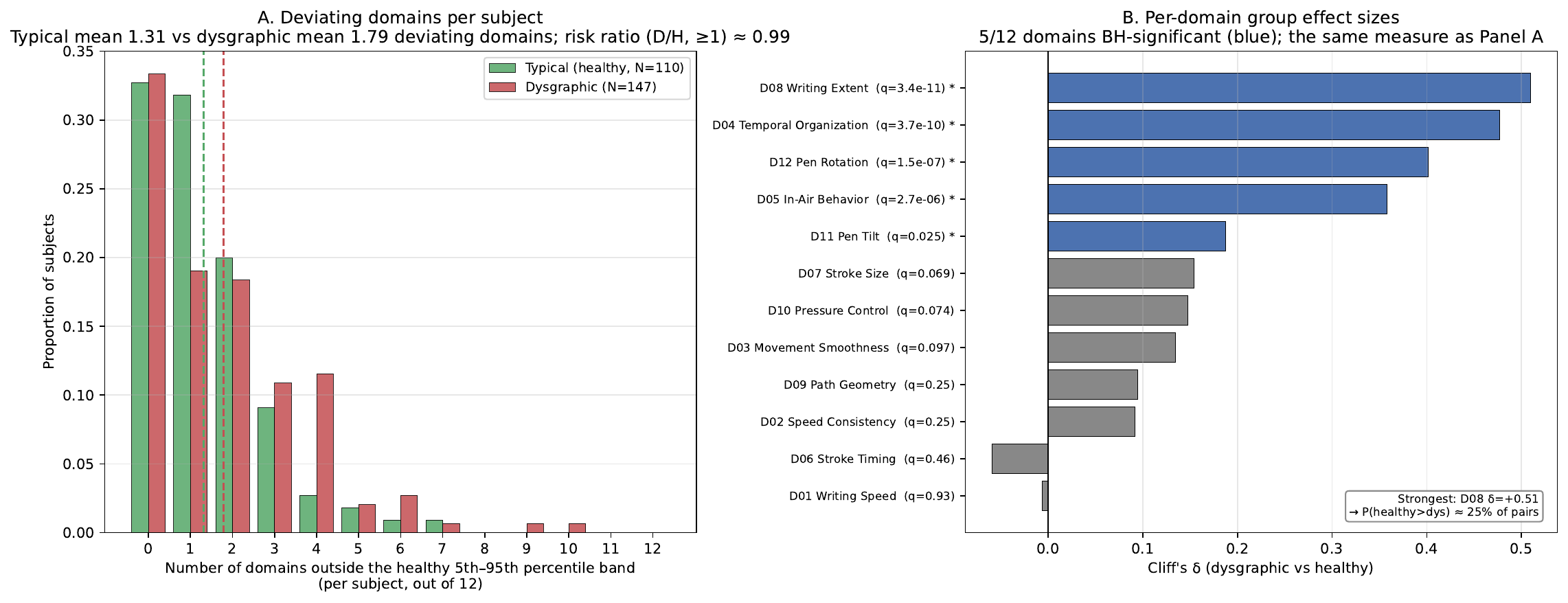}
\caption{Group-level effect is real, individual-level dichotomization is weak.
\textbf{(A)} Per-child count of the twelve domains on which the child's deviation
falls outside the healthy 5th--95th percentile band (the positional ``typical
range''). The typical and dysgraphic distributions overlap heavily: 67\% of
\emph{both} groups deviate on at least one domain (mean 1.31 vs.\ 1.79 domains;
dichotomized risk ratio at $\geq$1 domain $\approx0.99$), and separation appears
only in the tail ($\geq$3 domains: 15\% vs.\ 29\%). \textbf{(B)} Per-domain
group Cliff's $\delta$ for the same measure, with the five Benjamini--Hochberg-significant
domains in blue, the group-level signal that reproduces the known-groups analysis.
Even the strongest domain (D08 Writing Extent, $\delta=+0.51$) implies that a healthy
child scores higher than a dysgraphic child on roughly one in four cross-group pairs.
The same reference-based measure thus carries a real group signal (B) while producing
heavy individual overlap (A), the expected behavior of a normative-reference
instrument (positional, not diagnostic).}
\label{fig:s-groupvsind}
\end{figure*}

\end{appendices}